\documentclass[12pt,a4paper]{article}
\pdfoutput=1
 \usepackage{jheppub}
 \usepackage{amsmath}
 \usepackage{amsfonts}
 \usepackage{mathtools}
 \usepackage{amssymb}
 \usepackage{caption}
 \usepackage{subcaption}
 \usepackage{slashed}
\usepackage{float}
\usepackage{graphicx}
\graphicspath{ {project2 tex/} }

\def\n{\nu}

\newcommand{\be} {\begin{equation}}
\newcommand{\ee} {\end{equation}}
\newcommand{\bea} {\begin{eqnarray}}
\newcommand{\eea} {\end{eqnarray}}
\newcommand{\ba} {\begin{array}}
\newcommand{\ea} {\end{array}}
\newcommand{\nn} {\nonumber}

 \title{Gravitational couplings in 
 ${\cal N}=2$ string compactifications  and Mathieu Moonshine}
 \author{Aradhita Chattopadhyaya, Justin R. David}
\affiliation{Centre for High Energy Physics, Indian Institute of Science,\\
C. V. Raman Avenue, Bangalore 560012, India.}
\emailAdd{aradhita, justin@cts.iisc.ernet.in}

\abstract{
We evaluate the low energy gravitational 
couplings, $F_g$ in  the heterotic  $E_8\times E_8$ string theory 
compactified on  orbifolds of $K3\times T^2$ by $g'$ 
which acts as a  $\mathbb{Z}_N$ automorphism on 
$K3$  together with a $1/N$ shift along 
$T^2$. The orbifold   $g'$  corresponds to the  conjugacy 
classes of the Mathieu group $M_{24}$. 
The holomorphic piece of $F_g$ is given in terms of a 
polylogarithim with index $3-2g$ and 
predicts the Gopakumar-Vafa 
invariants in the corresponding dual type II 
Calabi-Yau compactifications. 
We show that  low lying Gopakumar-Vafa invariants for each of these
compactifications including the twisted sectors  are integers. 
We observe that the conifold singularity  for all these compactifications
occurs only when states in the twisted sectors become massless and 
the strength of the singularity is determined by the 
genus zero Gopakumar-Vafa invariant at this point in the moduli space. 
}

\begin{document}
\maketitle
\flushbottom

\section{Introduction}

The study of  the low energy  effective action of string compactifications
has proved to be important for various reasons.  
Higher derivative terms in the low energy effective action is relevant 
for evaluating sub-leading corrections to the Hawking-Bekenstein entropy
of black holes in these theories.   
The  one-loop low energy effective action encodes the quantum symmetries of 
the theory which are not visible from the classical effective action.
Usually  the one loop effective actions
for string compactifications are  automorphic functions in the moduli of the theory
and therefore encode the underlying symmetries of the theory. 
It often  easier to evaluate one-loop effective actions in 
one duality frame where it is exact and it leads to non-trivial predictions
in another frame. One simple example  which  has  these 
 features is the $R^2$ term in 
${\cal N}=4$  theories in $d=4$  which can be evaluated by a one-loop
computation in the type II frame, but yields predictions for an infinite sum of 
space time instanton effects due to  five branes in the heterotic frame 
\cite{Harvey:1996ir,Gregori:1997hi}.

Let us focus on the gravitational  coupling $F_g$ of the low energy 
effective action of ${\cal N}=2$ string theories in $d=4$. 
These couplings appear as the following terms in the effective action
\begin{equation}
S = \int F_g(y,  \bar y) F^{2g-2} R^2 ,
\end{equation}
where $F, R$ are the self dual part of the graviphoton and  the Riemann curvature. 
The coupling $F_g$ depends on the vector multiplets $y, \bar y$ of the theory. 
The canonical and well studied example of such a theory  is the $E_8\times E_8$ heterotic string 
theory compactified on $K3\times T^2$ with the standard embedding of the 
spin connection in to a $SU(2)$ of one of the $E_8$. 
Building on the earlier works 
\cite{Antoniadis:1992sa,Antoniadis:1992rq,Antoniadis:1993ze,Bershadsky:1993ta,Bershadsky:1993cx,Antoniadis:1995zn}, 
a detailed study of these couplings for this theory has been carried out 
by \cite{Harvey:1995fq} for $g=1$ and in 
 \cite{Marino:1998pg} for $g>1$ 
 by explicitly evaluating the one loop threshold integrals. 
This compactification is also the standard and well studied  example of 
${\cal N}=2$ string duality \cite{Kachru:1995wm,Klemm:1995tj}.  
The result from the heterotic side is one loop exact.  Moduli dependence
of the one loop threshold  corrections in this model have also  been studied earlier in
\cite{Mayr:1995rx,deWit:1995dmj,deWit:1996wq,LopesCardoso:1996nc,Stieberger:1998yi}. 
On the type II A side, the theory is compactified on 
the Calabi-Yau manifold   $X$ with Euler number $\chi(X)= -480$. 
In fact the Euler number of the dual Calabi-Yau   is predicted 
from  $F_g$. 

Though the full expression of $F_g$ is intricate, the 
 holomorphic\footnote{It is actually the anti-holomorphic part of 
 $F_g$ but we will take the complex conjugate and refer to it as the holomorphic part.}
  part of this coupling , 
 which can be obtained by sending $y\rightarrow \infty$
is simple and predicts certain topological invariants of the dual Calabi-Yau  three fold $X$. 
Lets call this holomorphic part $\bar F_g(y)$, the genus $g$ topological amplitude. 
The dual Calabi-Yau is  generally a 
$K3$ fibration  over a base $T^2$. The heterotic side corresponds to the semi-classical
limit when the volume of the base is large. 
The topological amplitude $\bar F_g(y)$  predicts the 
 the number of genus $g$  holomorphic curves in the $K3$ fibers of the dual
 Calabi-Yau $X$ in the semi-classical limit. 
The most convenient way to extract this information is using the 
observation of  Gopakumar and Vafa  who showed that 
a generating function for 
$\bar F_g(y)$  for any Calabi-Yau 
can be written in terms of integer invariants which are called
Gopakumar-Vafa invariants \cite{Gopakumar:1998ii,Gopakumar:1998jq}. 
Therefore the explicit calculation of $\bar F_g(y)$ from the heterotic side yields 
a prediction for the Gopakumar-Vafa invariants of the dual Calabi-Yau $X$
in the semi-classical limit. 

In this paper we generalize the evaluation of $F_g$ to the heterotic 
$E_8\times E_8$ string theory compactified on orbifolds of 
$K3\times T^2$. The orbifold $g'$ acts as a $\mathbb{Z}_N$ automorphism
on $K3$  together with the $1/N$ shift on one of the circles of $T^2$.
We consider the standard embedding in which the $SU(2)$ spin connection of 
the $K3$ is embedded in one of the $E_8$ of the theory. 
The orbifold $g'$  corresponds to the conjugacy classes of Mathieu
group $M_{24}$ listed in table  \ref{tt} and  they  all preserve ${\cal N}=2$ supersymmetry. 
 These compactifications were
introduced in \cite{Datta:2015hza,Chattopadhyaya:2016xpa} 
with the motivation of exploring the role of 
$M_{24}$ in string compactifications which was first studied in 
the original $K3\times T^2$ model by \cite{Cheng:2013kpa}.  
One of our goals in evaluating the $F_g$ for these orbifolds   is to 
determine the properties of the dual Calabi-Yau geometry 
on the type II A side and  to initiate a study of   how  the $M_{24}$ symmetry 
of $K3$  is seen in the Gopkumar-Vafa invariants  of the Calabi-Yau geometry. 
Recently the role of sporadic symmetry groups in the  elliptic genera  
of  Calabi-Yau 5-folds has been investigated \cite{Banlaki:2017vha}.

We now summarize our main  result for the holomorphic gravitational coupling $F_g$ for 
the orbifolds studied in this paper. 
Consider the twisted elliptic genus of $K3$ by
an automorphism  $g'$  of order $N$, which is defined as
\begin{eqnarray}
 F^{(r,s)} (\tau, z) = \frac{1}{N} 
 {\rm Tr}_{RR g^{\prime r} }
 [(-1)^{F_{K3} + \tilde{F}_{K3} } 
 g^{\prime s}
 e^{2\pi i z F_{K3}}  q^{L_0 - \frac{c}{24} } \bar q^{\tilde{L}_0 - \frac{\tilde{c}}{24}} ].
\end{eqnarray}
Here $F_{K3}, \tilde F_{K3}$ is left and the right moving  the fermion number. 
The trace is taken over the Ramond-Ramond sector twisted by $g^{\prime r}$. 
 The twisted elliptic genus can be written in the form
\begin{eqnarray}
F^{(0,0)} (\tau, z) &=& \alpha_{g'}^{(0, 0)} A(\tau, z) , \\ \nonumber
F^{(r, s)}(\tau, z) &=& \alpha_{g'}^{(r, s)} A(\tau, z)   + \beta_{g'}^{(r,s)} (\tau) B(\tau, z) , \\ \nonumber 
 & & \qquad r, s \in \{ 0, 1, \cdots N-1 \}  \;  \rm{with} \; (r, s) \neq (0, 0) 
\end{eqnarray}
where 
\begin{eqnarray}
 A(\tau, z) = \frac{\theta_2^2(\tau, z) }{\theta_2^2(\tau, 0)} + 
 \frac{\theta_3^2(\tau, z) }{\theta_3^2(\tau, 0)} +
 \frac{\theta_4^2(\tau, z) }{\theta_4^2(\tau, 0)}, \qquad
 B(\tau, z) = \frac{\theta_1^2(\tau, z) }{\eta^6(\tau) }
\end{eqnarray}
are Jacobi forms which transform under $SL(2, \mathbb{Z})$ with index 1 and 
weight $0$ and $-2$ respectively. 
$\alpha_{g'}^{(r, s)}$ are numerical constants and $\beta_{g'}^{(r, s)}$ is a 
weight 2 modular form under $\Gamma_0(N)$.   
After the discovery of the Mathieu moonshine symmetry in the 
elliptic genus of $K3$ \cite{Eguchi:2010ej}, 
the twining character
$F^{(0,1)}$ for all the $M_{24}$ conjugacy classes was  first found in \cite{Cheng:2010pq,Gaberdiel:2010ch,Eguchi:2010fg}. 
For  $g^{\prime}$  given in table \ref{tt}, the corresponding to the  full twisted elliptic genus 
$M_{24}$ have been explicitly determined  in 
\cite{Chattopadhyaya:2017ews,Gaberdiel:2012gf}. We list them in appendix \ref{frslist} 
for completeness. 
Now given the twisted elliptic genus, consider the following 
weight $2g$ quasi-modular form  under $\Gamma_0(N)$
\begin{eqnarray}\label{defcrsfrs0}
f^{(r,s)}(\tau) {\cal P}_{2g} ( G_2,  G_4, G_6,  \cdots , G_{2g}) 
&=& \sum_{l \in \frac{\mathbb{Z} }{N}} c_{g-1}^{(r,s)} (l, 0) q^l , \\ \nonumber
\hbox{where}\qquad f^{(r,s)} (\tau) &=&  \frac{1}{2\eta^{24} (\tau) } 
 E_4 
 \left[ \frac{1}{4} \alpha_{g'}^{(r,s)} E_6  - 
 \beta_{g'}^{(r,s)} (\tau) E_4 \right], \\ \nonumber
 G_{2k} &=&  2\zeta(2k) E_{2k}.
\end{eqnarray}
 $E_{2k}$ are Eisenstein series of weight $2k$ and $\zeta$ refers to the zeta function. 
${\cal P}_{2g}$ is related to the Schur polynomial ${\cal S}$  of order $g$ by 
\begin{equation}
{\cal P}_{2g} ( x_1, x_2, \cdots x_g) = - {\cal S} ( x_1, \frac{1}{2} x_2, \cdots \frac{1}{g} x_g) .
\end{equation}
Then the topological amplitude $\bar F_g$ is given by 
\bea\label{fhol0}\nonumber
\bar{F}_g^{{\rm hol}} (y) =\frac{(-1)^{g-1}}{\pi^2} \sum_{s=0}^{N-1} \left( \sum_{m>0}e^{-2\pi i n_2s/N}c^{(r,s)}_{g-1}(m^2/2,0){\rm Li}_{3-2g}(e^{2\pi im\cdot y})+\frac{1}{2}c_{g-1}^{(0,s)}(0,0)\zeta(3-2g)
\right) .\\
\eea
The sum over lattice points $m>0$ refers to  the following lattice points  $(n_1, n_2)$, 
$n_1 \in \frac{\mathbb{Z}}{N}, n_2\in \mathbb{Z}$ with the restrictions
\begin{eqnarray}
n_1, n_2 \geq 0, \quad\hbox{but } ( n_1, n_2 ) \neq ( 0, 0 ), \\ \nonumber
  ( r/N, - n_2) , \qquad \hbox{with}\;\;  n_2>0 \;\;\hbox{and}\;\; r n_2 \leq N.
\end{eqnarray}
 $y = (T, U)$ is the K\"{a}hler and complex structure of the torus $T^2$, $m^2= 2n_1n_2$ and
 $m\cdot y =  n_1 T + n_2 U$. 
 The functions ${\rm Li}_{3-2g}$ are polylogarithm functions of order $3-2g$. 

From  (\ref{defcrsfrs0}) we see that indeed it is the  
coefficients of the twisted elliptic genus 
of $K3$ which 
forms the basic input data for the topological amplitude $\bar F_g(y)$. 
It is   in this  sense $M_{24}$ symmetry of  $K3$ is carried over
to the topological amplitude.   This is a generalization of the observation 
by \cite{Cheng:2013kpa} in which the elliptic genus  of $K3$ which determines the factor $E_4 E_6/\eta^{24}$, 
is the crucial input data for the topological amplitude for the unorbifolded model. 
Now comparing the instanton contributions in 
$\bar F_g(y)$ with the form of the topological amplitude
written in terms of the Gopakumar-Vafa invariants we can extract out 
these invariants. It is  apriori  not clear that  the invariants
will be integers since the coefficients $c_{g-1}^{(r,s)} $
in (\ref{defcrsfrs0}),  themselves are not integers. 
However to the level we have verified the Gopakumar-Vafa invariants are all integral.
This forms a  simple consistency check of our result. 
In fact we will see that  once the genus zero Gopakumar-Vafa invariants are integers, 
the higher genus invariants are assured to be integers. This is shown for  genus $g=1, 2, 3$. 
The constant term in (\ref{fhol0}) contains the information of the 
Euler character of the dual Calabi-Yau X. 
The Euler character of the Calabi-Yau dual to all the orbifolds considered 
in this paper is listed in table \ref{chi}. 

We then study the conifold singularities of the $\bar F^{\rm hol}_g$ which correspond 
to points in moduli space  of enhanced gauge symmetry. 
We observe that there are no conifold singularities from the untwisted sector and all 
conifold singularities  are due to  twisted sector states becoming massless. 
The strength of this singularity is proportional to the genus zero Gopakumar-Vafa
invariant corresponding to this state. The list of low lying genus zero Gopakumar-Vafa invariants
for $g'$ which we refer to as CHL orbifolds\footnote{These are geometric actions, which 
keep the Hodge number of $K3$;  $h_{1,1}\geq 1$. }  are provided in appendix \ref{GVZ}.
For the rest of the $g'$ in table \ref{tt}. 
we provide an ancillary Mathematica code from which the genus zero  Gopakumar-Vafa invariants
can be evaluated and seen to be integral.

We briefly mention  the method we adopt to evaluate  $F_g$. 
In the heterotic frame it is given in terms of a one loop integral over the fundamental 
domain.  We basically follow the  method of orbits adopted in many works
in this subject starting from \cite{Dixon:1990pc}.
See \cite{Florakis:2016boz} for a recent discussion
of these methods. 
However  we do not directly apply the lattice reduction theorem of 
Borcherds \cite{Borcherds:1996uda} as done by earlier works 
\cite{Marino:1998pg,Klemm:2005pd,Weiss:2007tk} for evaluating $F_g,\; g>1$. 
This is because the application of the reduction theorem when the 
integrand has modular forms of $\Gamma_0(N)$ is rather intricate 
and we find it  is easier to proceed directly and carry out each of the 
steps involved in the integrations.  Integrals involving modular forms 
of $\Gamma_0(N)$ occurring in the theta lift of the twisted elliptic genus of 
$K3$ were first done in \cite{David:2006ji} and \cite{David:2006ud}.

The organization of the paper is as follows. 
In  section \ref{review},  we review aspects of the  $\mathbb{Z}_N$ 
orbifold compactifications 
of the   $E_8\times E_8$ heterotic theory $K3\times T^2$ we study. In particular 
we will present the new supersymmetric index which forms the basic 
ingredient for the gravitational threshold integral.  Next in section 
\ref{setupint} we write down the expressions for the coupling $F_g$ 
for compactifications on orbifolds of $K3\times T^2$ . 
In section \ref{Intgt} , as a warm up we  first evaluate the 
gravitational coupling $F_1$. 
We do this because the integrand as well as 
performing the integral is a straight forward generalization
of the one done for the $K3\times T^2$ theory by \cite{Harvey:1995fq} 
and provides checks for
our calculation. 
 Next we  present the result for the gravitational coupling $F_g(y, \bar y)$ for $g>1$ 
and  extract the holomorphic coupling $\bar F_g^{\rm hol} (y)$. 
In section \ref{GVCS} we study the properties of 
the genus $g$ topological amplitude and observe that the 
Gopakumar-Vafa invariants of the dual Calabi-Yau 
 predicted by our calculation.  We then study the conifold singularities
 of $\bar F_g^{\rm hol} (y)$. 
In section \ref{conclusion}  we  present our conclusions. 
The appendix  \ref{modt} contains the modular properties of the integrand
involved in evaluating $F_g(y, \bar y)$. 
Appendices  \ref{detailg1}, \ref{detailg2}    contain the details of the integrations to obtain the 
gravitational thresholds. 
Appendix \ref{GVZ} lists the Gopakumar-Vafa invariants of 
for all $g'$ corresponding to CHL orbifolds. 
Finally appendix \ref{frslist}  lists the elliptic genera of all the orbifolds considered
in the paper for completeness.

\begin{table}[H]
	\renewcommand{\arraystretch}{0.5}
	\begin{center}
		\vspace{0.5cm}
				\begin{tabular}{|c|c|}
					\hline
					&  \\
					Conjucay Class & Order \\
					\hline
					&  \\
					1A & 1\\
					2A & 2\\
					3A & 3\\
					5A & 5\\
					 7A & 7\\
					 11A & 11\\
					 23A/B & 23 \\
					\hline
					 & \\
					 4B & 4\\
					 6A & 6 \\
					 8A & 8 \\
					 14A/B & 14\\
					 15A/B & 15\\
					\hline
					 & \\
					 2B & 2 \\
					 3B & 3 \\ 
					\hline
				\end{tabular}
		\end{center}
	\vspace{-0.5cm}
\caption{{\footnotesize{  
Conjugacy classes of $M_{24}$ studied in the paper.\\
The twisted genera of classes  2B and 3B  are such that the twining genera matches with the $F^{(0,1)}$ listed in \cite{Eguchi:2010ej} .  
These  orbifolds are of order 4 and 9 respectively in our analysis.
We refer to the classes 2A, 3A, 5A, 7A, 4B, 6A, 8A  as the CHL orbifolds.}}} \label{tt}
\renewcommand{\arraystretch}{0.5}
\end{table}

\section{Heterotic string on orbifolds of $K3\times T^2$} \label{review}

In this section we briefly describe the general class of ${\cal N}=2$ compactifications 
we  will study. 
Consider the $E_8\times E_8$ heterotic string theory compactified 
on $K3\times T^2$ 
in which the $SU(2)$  spin connection of $K3$ is embedded in 
one of the $E_8$'s which is  called the standard embedding. 
We then orbifold  by  a freely acting $\mathbb{Z}_N$ which 
acts as a $g'$ automorphism on $K3$ together with a 
$1/N$ shift along one of the circles of $K3$. 
The $g'$ action corresponds to any of the $26$ conjugacy classes
of Mathieu group listed in the table.
In 
the standard embedding one of the $E_8$ lattice breaks to a 
$D6\otimes D2$ lattice. The $SU(2)$ spin connection of 
$K3$  couples to the fermions of $D2$ lattice. 
These left moving fermions with the left moving bosons of the $K3$ 
combines with the right moving supersymmetric $K3$ CFT to form 
a $(6, 6)$  conformal field theory. 
The internal conformal field theory is  given by 
\begin{equation}
 {\cal H}^{\rm internal} = {\cal H}_{D2K3}^{(6,6)} \otimes {\cal H}_{D6}^{(6,0)}
 \otimes {\cal H}_{E_8}^{(8,0)} \otimes {\cal H}_{T^2}^{(2,3)} 
\end{equation}
The  $g'$ orbifold  acts as a $\mathbb{Z}_N$ automorphism on the 
$(6,6)$ CFT which preserves its 
$SU(2)$ R-symmetry together with a $1/N$ shift on one of the circles 
of the $T^2$ CFT. 
Thus these compactifications preserves ${\cal N}=2$ supersymmetry 
in $4$ dimensions.  We refer to these  theories as compactifications of 
the $E_8\times E_8$ theory on the orbifold
$K3\times T^2/\mathbb{Z}_N$. 

The  starting point in evaluating both gauge and gravitational  threshold corrections is to 
obtain the 
new supersymmetric index  of the internal CFT. 
This is defined by 
\begin{eqnarray}\label{newind}
 {\cal Z}_{\rm new} = \frac{1}{\eta^2 (\tau) } 
 {\rm Tr}_R[ (-1)^{F} F q^{L_0 - \frac{c}{24} } \bar q^{\tilde{L}_0 - \frac{\tilde{c}}{24} }
 ].
\end{eqnarray}
Here $F$ refers to the right moving fermions number which is given by 
sum of the right moving fermion number of the $T^2$ CFT together with the 
right moving fermion number of the $K3$ CFT
\begin{equation}
 F= F^{T^2} + F^{K3}. 
\end{equation}
The trace in (\ref{newind}) is taken over the Ramond sector of the right moving 
supersymmetric  internal conformal field theory.
Note that $(c, \tilde c ) = ( 22, 9)$. 

In \cite{Chattopadhyaya:2016xpa},  the new supersymmetric index for compactifications on 
the orbifolds $K3\times T^2/\mathbb{Z}_N$ was evaluated and 
it was shown that it can be written in terms of the twisted 
elliptic genus of $K3$. Let us  briefly review the result. 
First we define the twisted elliptic genus of $K3$ which is given by 
\begin{eqnarray}
 F^{(r,s)} (\tau, z) = \frac{1}{N} 
 {\rm Tr}_{RR g^{\prime r} }
 [(-1)^{F_{K3} + \tilde{F}_{K3} } 
 g^{\prime s}
 e^{2\pi i z F_{K3}}  q^{L_0 - \frac{c}{24} } \bar q^{\tilde{L}_0 - \frac{\tilde{c}}{24}} ].
\end{eqnarray}
Here $g'$ is the $\mathbb{Z}_N$ automorphism on $K3$ corresponding to the 
conjugacy classes listed in table \ref{tt}. However the construction goes through for any consistent orbifold on $K3$.
The trace is taken in the Ramond-Ramond sector which is twisted by $g^{\prime r}$. 
The twisted elliptic genus can be written in the general form 
\begin{eqnarray} \label{genform}
 F^{(0,0)} (\tau, z) &=& \alpha_{g'}^{(0, 0)} A(\tau, z) , \\ \nonumber
 F^{(r, s)}(\tau, z) &=& \alpha_{g'}^{(r, s)} A(\tau, z)   + \beta_{g'}^{(r,s)} (\tau) B(\tau, z) , \\ \nonumber 
 & & \qquad r, s \in \{ 0, 1, \cdots N-1 \}  \;  \rm{with} \; (r, s) \neq (0, 0) 
\end{eqnarray}
where 
\begin{eqnarray}
 A(\tau, z) = \frac{\theta_2^2(\tau, z) }{\theta_2^2(\tau, 0)} + 
 \frac{\theta_3^2(\tau, z) }{\theta_3^2(\tau, 0)} +
 \frac{\theta_4^2(\tau, z) }{\theta_4^2(\tau, 0)}, \qquad
 B(\tau, z) = \frac{\theta_1^2(\tau, z) }{\eta^6(\tau) }
\end{eqnarray}
are Jacobi forms which transform under $SL(2, \mathbb{Z})$ with index 1 and 
weight $0$ and $-2$ respectively. 
$\alpha_{g'}^{(r, s)}$ are numerical constants and $\beta_{g'}^{(r, s)}$ is a 
weight 2 modular form under $\Gamma_0(N)$. 
For example for $g' \in pA$ with $p =2,  3, 5, 7$ we have 
\begin{eqnarray}
 \alpha^{(0,0)}_{pA} &=& \frac{8}{p}, \qquad \alpha^{(r,s)}_{pA} = \frac{8}{p( p+1)},\; (r,s) \neq (0, 0)  \\ \nonumber
 \beta^{(0, s)}(\tau) &=& - \frac{2}{p+1} {\cal E}_p(\tau), \;\;  {\rm {for}}\; 1\leq s\leq p -1 \\ \nonumber
 \beta^{(r, rk)} (\tau)  &=& \frac{2}{p(p+1)}{\cal E}_p(\frac{\tau +k}{p} ), \\ \nonumber
 && \; {\rm{for}}   1\leq r \leq p-1, \;  1\leq k \leq p-1
\end{eqnarray}
Here 
\begin{equation}
 {\cal E}_N = \frac{12i}{\pi (p-1)} \partial_\tau[\ln\eta(\tau) - \ln \eta(N\tau) ]
\end{equation}
is a modular from which transforms with weight $2$ under $\Gamma_0(N)$. The 
twisted elliptic genus transforms under $SL(2, \mathbb{Z})$ as 
\begin{eqnarray}\label{twsttrans}
F^{(r, s)} \left( \frac{a\tau + b}{c\tau + d} , \frac{z}{c\tau + d} \right) 
= \exp\left( 2\pi i \frac{c z^2}{ c\tau + d} \right) 
F^{( cs + ar , ds + b r) } ( \tau, z) 
\end{eqnarray}
for 
\begin{equation} \label{sl2z}
a, b, c, d \in \mathbb{Z}, \qquad ad-bc = 1
\end{equation}
The indices in (\ref{twsttrans})   $cs + ar$ and $ds + br$ are taken to be mod $N$.

Appendix \ref{frslist}   lists the  twisted elliptic genus $F^{(r, s)}(\tau, z) $ for 
corresponding to the conjugacy classes $g'$ in table. 
Given the twisted elliptic genus,  we an read out 
$\alpha^{(r,s)}_{g'}, \beta^{(r,s)}_{g'}$   using the general form 
given in (\ref{genform}). 
Then the new supersymmetric index  for the standard embedding compactifications
of the $E_8\times E_8$ heterotic string on the orbifold $K3\times T^2/\mathbb{Z}_N$ 
is given by 
\begin{equation} \label{resnews}
 {\cal Z}_{\rm new}(q, \bar q) = - 2 \frac{1}{\eta^{24} (\tau)} 
 \Gamma_{2,2}^{(r, s)} E_4 
 \left[ \frac{1}{4} \alpha_{g'}^{(r,s)} E_6  - 
 \beta_{g'}^{(r,s)} (\tau) E_4 \right].
\end{equation}
Here it is understood that the indices $r, s$ are summed from $0$ to $N-1$. 
In (\ref{resnews}),  $\Gamma_{2,2}^{(r,s)}$ is the lattice sum on $T^2$ which is defined as
\begin{eqnarray} \label{gamma22}
 \Gamma_{2,2}^{(r,s)} (q, \bar q) 
 &=& \sum_{\stackrel{m_1, m_2, n_2 \in \mathbb{Z} }{n_1= \mathbb{Z} + \frac{r}{N} } }
 q^\frac{p_L^2}{2} \bar q ^{\frac{p_R^2}{2}} e^{2\pi i m_1 s /N}, 
 \\ \nonumber
 \frac{1}{2} p_R^2 &=& 
 \frac{1}{2T_2 U_2} |-m_1 U + m_2 + n_1 T + n_2 TU |^2 , \\ \nonumber
 \frac{1}{2}p_L^2 &=& \frac{1}{2} p_R^2 + m_1n_1 + m_2 n_2
\end{eqnarray}
and $T, U$ are the K\"{a}hler and complex structure of the torus $T^2$. 
It is also understood in (\ref{resnews}), that there is a sum over the lattice momenta
$m_1, m_2$ as well as the winding numbers $n_1, n_2$. 
Note that in  (\ref{resnews}) its only the lattice sum that depends on both $q$ and $\bar q$, 
the Eisenstein series $E_4, E_6$  as well 
as the $\Gamma_0(N)$ weight $2$ form $\beta_{g'}^{(r, s)}$ 
depends only on the holomorphic co-ordinate $q$. 
It is instructive to examine the situation in which there is no orbifold performed. 
Then $p=1$ and 
 $\alpha_{1A} = 8$
and there are no twisted sectors. 
Thus  the new supersymmetric index reduces to 
\begin{equation}
 {\cal Z}_{\rm new}(q, \bar q) = - 4 \frac{E_4(\tau) E_6(\tau) }{\eta^{24} (\tau)} 
 \Gamma_{2,2} (q, \bar q ) 
\end{equation}
Here $\Gamma_{2,2}$ is the usual lattice sum without phases and twists. 
For later purpose we define the following $\Gamma_0(N)$ form, which occurs 
in the new supersymmetric index given in (\ref{resnews}). 
\begin{eqnarray}\label{defcrsfrs}
f^{(r,s)} (\tau) &=&  \frac{1}{2\eta^{24} (\tau) } 
 E_4 
 \left[ \frac{1}{4} \alpha_{g'}^{(r,s)} E_6  - 
 \beta_{g'}^{(r,s)} (\tau) E_4 \right], \\ \nonumber
 &=&\sum_{ l \in \frac{\mathbb{Z}}{N} }  c^{(r, s)} ( l) q^{l}.
\end{eqnarray}
Using the fact that $\beta_{g'}^{(r, s)} ( \tau) $ is a $\Gamma_0(N)$ form and 
(\ref{twsttrans}) it is easy to 
see that the $f^{(r, s)}(\tau)$ has the following transformation 
property under  the  $SL(2, \mathbb{Z})$ generators
\begin{eqnarray} \label{modtrans}
& & f^{(r, s) } ( \tau + 1 ) =  f^{(r, s + r) } ( \tau) , \\ \nonumber
& &f^{(r, s) } ( -\frac{1}{\tau} ) =  ( - i \tau) ^{-2} ) f^{(N-r, s ) } (\tau)  = 
( -i \tau)^ {-2} f^{ ( r, N-s) } ( \tau) .
\end{eqnarray} 

Given the new supersymmetric index we can read out the difference between the 
number of vectors and hypers in the spectrum \cite{Harvey:1995fq}. 
This is given by 
\begin{eqnarray}
N_h - N_v = - \sum_{s =0}^{N-1} c^{(0, s)}(0) .
\end{eqnarray}
Here, we have also  included  the four  $U(1)$'s  
resulting from the metric and the Neveu-Schwarz $B$-field with 
one index along the $T^2$ in the counting of the vectors. 
This data also gives the Euler number of the corresponding 
 type II dual  Calabi-Yau compactification which is given by 
 \begin{equation}
 \chi (X) = - 2 ( N_h - N_v) = 2 \sum_{s =0}^{N-1} c^{(r, s)}(0) .
 \end{equation}
 For the compactification on $K3\times T^2$ the value of 
 $N_h-N_v = 240$ and therefore the corresponding Euler number of the 
 dual $K3$-fibered Calabi-Yau manifold is $\chi(X) = - 480$. 
 These data for all the orbifolds considered in this paper 
 is provided in table \ref{chi}.

In the subsequent sections we will use the new supersymmetric index in (\ref{resnews}) 
as a starting point  and evaluate the gravitational one-loop corrections 
for heterotic compactifications on orbifolds $K3\times T^2/\mathbb{Z}_N$.

\section{The integral for gravitational thresholds} \label{setupint}

It is first easy to discuss the  one loop integral which captures the gravitational 
correction with no graviphoton insertions.  The result  of this integral provides the moduli dependence 
of the following higher derivative term in the ${\cal N}=2$ effective action 
for these compactifications
\begin{equation}
 I_1 (T, U)  = R_{+}^2 F_1(T, U) .
\end{equation}
Here $R_+$ refers to the anti-self-dual Riemann tensor.  
From  the  earlier works of \cite{Antoniadis:1992sa,Antoniadis:1992rq},  a  nice compact
 expression for the $F_1$  has been obtained in \cite{Harvey:1995fq}. This 
 is given by \footnote{Note that we have  normalized $F_1$  so that 
  the result for the integral agrees with \cite{Harvey:1995fq} for the
   $K3\times T^2$ compactification for $N=1$.}
\begin{eqnarray} \label{f1def}
 F_1 (T, U)  = -\frac{1}{4} \int_{\cal F}  \frac{d^2 \tau}{\tau_2} \left\{ 
  \frac{1}{\eta^2 (\tau) } 
 {\rm Tr}_R[ (-1)^{F} F q^{L_0 - \frac{c}{24} } \bar q^{\tilde{L}_0 - \frac{\tilde{c}}{24} }
 ]\left[ 
 E_2 - \frac{3}{\pi \tau_2} \right] - b_{\rm{grav}} \right\}.
\end{eqnarray}
The subtraction by $b_{grav}$ is to ensure that the integral is well 
defined for $\tau_2 \rightarrow \infty$. 
Note that the trace over the internal conformal field theory is 
precisely what occurs in the new supersymmetric index  given in (\ref{newind}).  
Therefore to obtain the moduli dependence of the gravitational correction for the 
orbifold compactifications in this paper, we substitute 
the result for the new supersymmetric index  (\ref{resnews}) in (\ref{f1def})
This results in 
\begin{eqnarray} \label{g1int}
 F_1(T, U)   =  \frac{1}{2} \int \frac{d^2\tau}{\tau_2}  \frac{1}{\eta^{24} (\tau)} 
 \Gamma_{2,2}^{(r, s)} E_4 
 \left[ \frac{1}{4} \alpha_{g'}^{(r,s)} E_6  - 
 \beta_{g'}^{(r,s)} (\tau) E_4 \right]\left[ 
 E_2 - \frac{3}{\pi \tau_2} \right],
\end{eqnarray}
where we have ignored the constant necessary to regulate the integral. 
From the properties (\ref{latticemod}) and (\ref{modtrans}) we see that the integrand 
is invariant under $SL(2, \mathbb{Z})$ transformations. 
The next step is to perform the integral over the fundamental domain which is done in 
 section \ref{genusone}.

Let us now discuss the integral which results in the moduli dependence of the 
following term in the one loop effective action
\begin{equation}
 I_ g (T, U) =  R_+^2 F_+^{2g - 2} F_g (T, U), 
\end{equation}
where $F_+$ is the anti-self dual field strength of the graviphoton and $g>1$. 
From the  world sheet analysis of \cite{Antoniadis:1995zn} we see that the one loop integral is given by 
\begin{eqnarray} \label{fgexp}
F_g &=& \frac{1}{ 2\pi^2( g!) ^2} \int \frac{ d^2 \tau}{\tau_2} \left\{  \frac{1}{\tau_2^2 \eta^2(\tau)}
{\rm Tr}_R
\left[   ( i \bar \partial X)^ {( 2g - 2) } 
(-1)^{F} F q^{L_0 - \frac{c}{24} } \bar q^{\tilde{L}_0 - \frac{\tilde{c}}{24} } \right ]  \right. \\ \nonumber
&& \qquad \qquad \qquad  \left.  \langle \prod_{i=1}^g \int d^2 x_i Z^1 \partial Z^2 (x_i)  
\prod_{j = 1}^g \int d^2 y_j \bar Z^2  \partial \bar Z^1 ( y_j)  \rangle  \right\}.
\end{eqnarray}
Here $X$ is the complex co-ordinate on the $T^2$ and $Z^1, Z^2$ and the complex
co-ordinates of the transverse non-compact space time bosons. 
The above result is obtained by carrying out the Wick contractions of the vertex operators for the 
graviton and the graviphoton insertions as mentioned in \cite{Antoniadis:1995zn}. 
Note that the correlators of the space time bosons and the internal 
conformal field theory factorize. 
The $2g-2$ insertions of $\bar \partial X$  arise from the $(2g-3)$ graviphoton vertices in the 
$(0)$-ghost picture and one $\bar \partial X$  that appears in the picture changing operator. 
Since the correlator is evaluated on the torus, the  insertions of $\bar \partial X$ 
can 
contribute only through the zero modes.
Therefore this   can be replaced  as follows
\begin{equation}\label{prexp}
 ( i \bar\partial X)^{( 2g - 2)}   \rightarrow \left(  \frac{(p_R^{(r, s)} ) } {\sqrt{2T_2 U_2} }
 \right)^{(2g -2) }.
\end{equation}
The super script $(r, s)$ for $p_R$ indicates the sector over which the right 
moving momentum belongs to.  When $r\neq 0$ the winding 
number $n_1$ on one of the circles of the 
$T^2$ is quantized in units of $\mathbb{Z} + \frac{r}{N}$ since 
the right moving momentum belongs to the  $r$th twisted sector of the $1/N$ shift action.  
Then trace over the internal conformal field theory can be written as
\footnote{In the analysis of  \cite{Antoniadis:1995zn}, the trace over the internal conformal field 
theory was referred to as $C_\epsilon(\bar\tau)$.  Further more, what we call as left movers 
is right movers in \cite{Antoniadis:1995zn} and vice-versa. We follow the notations of 
\cite{Marino:1998pg}.}
\begin{eqnarray} \label{intrace}
&& \frac{1}{\eta^2(\tau) } {\rm Tr}_R
\left[   ( i \partial X)^{( 2g - 2)}  
(-1)^{F} F q^{L_0 - \frac{c}{24} } \bar q^{\tilde{L}_0 - \frac{\tilde{c}}{24} } \right ]
= \\ \nonumber 
& & \qquad\qquad\qquad\quad 2 \frac{1}{\eta^{24} (\tau)} 
 \Gamma_{2,2}^{(r, s)} \left(  \frac{(p_R^{(r, s)} ) } {\sqrt{2T_2 U_2} }
 \right)^{(2g -2) }  E_4 
 \left[ \frac{1}{4} \alpha_{g'}^{(r,s)} E_6  - 
 \beta_{g'}^{(r,s)} (\tau) E_4 \right].
\end{eqnarray}
This trace is evaluated using the input in (\ref{prexp}) and repeating the 
steps which led to the result in (\ref{resnews}).  The only difference is the 
insertions of the lattice momenta $p_R^{(r,s)}$. 
Note that the sectors $(r, s)$ are summed over, and it is always understood that
the lattice momenta and winding are summed. 

Now we need to simplify the correlators over the free non-compact bosons. 
To this end we follow \cite{Antoniadis:1995zn} and  use the generating function
for these correlators 
\begin{equation}
G( \lambda, \tau, \bar \tau)  = 
\sum_{g=1}^\infty \frac{1}{(g!)^2} ( \frac{\lambda}{\tau_2}  )^{2g}
\langle \prod_{i=1}^g 
\int d^2 x_i Z^1 \partial Z^2 (x_i)  
\prod_{j = 1}^g \int d^2 y_j \bar Z^2  \partial \bar Z^1 ( y_j)  \rangle .
\end{equation}
The correlation functions are normalized free field correlators of the 
space-time bosons. This generating function is given by 
\begin{equation} \label{genfn}
 G(\lambda,  \tau, \bar \tau) = \left( \frac{2\pi i \lambda \eta^3}{\theta_1 ( \lambda, \tau) } \right) ^2 
 e^{ - \frac{\pi \lambda^2}{\tau_2} } .
\end{equation}
To use this generating function for the correlators, we consider
\begin{equation}
 F(\lambda, T, U) = \sum_{g=1}^\infty \lambda^{2g} F_g( T, U) .
\end{equation}
Then using the result  (\ref{intrace}) and  (\ref{genfn})  in
(\ref{fgexp}), the one loop amplitude simplifies to 
\begin{equation} \label{fg2}
 F(\lambda, T, U) = \frac{1}{\pi^2} \int \frac{d^2\tau}{\tau_2} 
  \frac{1}{\eta^{24} (\tau)} 
 \Gamma_{2,2}^{(r, s)}
 E_4 
 \left[ \frac{1}{4} \alpha_{g'}^{(r,s)} E_6  - 
 \beta_{g'}^{(r,s)} (\tau) E_4 \right]
 \left[ 
  \left( \frac{2\pi i \lambda \eta^3}{\theta_1 ( \tilde  \lambda, \tau) } \right) ^2 
 e^{ - \frac{\pi \tilde \lambda^2}{\tau_2} }  \right]^{(r, s)}.
\end{equation}
where 
\begin{equation}
 \tilde \lambda = \frac{p_R^{(r, s)} \lambda}{\sqrt{2 T_2 U_2}}.
\end{equation}

To perform this integral we follow the approach in \cite{Marino:1998pg}. 
The reciprocal of the theta function admits  the expansion
\begin{equation}\label{thetexp}
 \left ( \frac{2\pi  \tilde \lambda \eta^3}{\theta_1 (\tilde \lambda,  \tau) } 
 \right)^2 e^{- \frac{\pi \tilde \lambda^2}{\tau_2} } 
 = \sum_{k=0}^\infty \tilde \lambda^{2k} {\cal P}_{2k} ( \hat G_2, \ldots, G_{2k}) ,
\end{equation}
where   ${\cal P}_{2k}$ is  a polynomial   related to the Schur polynomial by 
\begin{eqnarray}
 {\cal P}_{2k} ( \hat G_2, \ldots, , G_{2k} ) = - {\cal S}_k \left( \hat G_2, \frac{1}{2} G_4, \frac{1}{3} G_6, 
 \ldots , \frac{1}{k} G_{2k} \right),  
 \end{eqnarray}
 where the Schur polynomials are in turn defined by the expansion
 \begin{eqnarray}
 && \exp\left[ \sum_{k=1}^\infty x_k z^k \right] = \sum_{k =0}^\infty {\cal S}_k ( x_1, \cdots x_k ) z^k , 
 \\ \nonumber
&& {\cal S}_1(x_1) =  x_1, \qquad {\cal S}_2 (x_1, x_2) = \frac{x_1^2}{2} + x_2, \qquad
 {\cal S}_2(x_1, x_2, x_3) = \frac{x_1^3}{6} + x_1 x_2 + x_3.
\end{eqnarray}
Using these definitions we have 
\begin{eqnarray}
 & &{\cal P}_2 ( \hat G_2) = - \hat G_2, \quad  
 {\cal P}_4 ( \hat G_2, G_4) = - \frac{1}{2} ( \hat G_2^2 + G_4),  \\ \nonumber
 & & {\cal P}_3 ( \hat G_2, G_4, G_6) = - \frac{1}{6}(  \hat G_2^3  + \hat G_2 \hat G_4  ) - \frac{1}{3} G_6 .
\end{eqnarray}
where the $G$'s are  normalized Eisenstein series given by 
\begin{equation}
 G_{2k} = 2 \zeta(2k) E_{2k}, \qquad \hat E_{2}(\tau)  = E_2 - \frac{3}{\pi \tau_2}.
\end{equation}
Note that ${\cal P}$ is holomorphic, except for the occurrence of $\hat G_2$ and it 
transforms as  a modular form of weight $2k$.
Substituting the expansion (\ref{thetexp}) in  (\ref{fg2}), we obtain 
\begin{eqnarray}
 F(\lambda, T, U) = \sum_{k=0}^\infty 
 \frac{\lambda^{2 (k+1)}}{\pi^2 ( 2 T_2 U_2 )^k } 
 \int_{{\cal F}} \frac{d^2\tau}{\tau_2} \tau_2^{2k} 
  \frac{1}{\eta^{24} (\tau)} 
 \Gamma_{2,2}^{(r, s)}
 E_4 
 \left[ \frac{1}{4} \alpha_{g'}^{(r,s)} E_6  - 
 \beta_{g'}^{(r,s)} (\tau) E_4 \right] ( p_R^{(r,s)} )^{2k} {\cal P}_{2k +2}.\nonumber \\
\end{eqnarray}
Note that the integrand is an invariant under $SL(2, \mathbb{Z})$. 
This can be seen using the properties in (\ref{prmodtrans}) and (\ref{modtrans})

\section{Evaluating  the  integral for gravitational thresholds } \label{Intgt}

\subsection{$g=1$}\label{genusone}

We will first perform the integral in (\ref{g1int}). As it will be clear 
subsequently,  it is simpler to treat the case of $g=1$ separately from the  integral 
for $g>1$. 
We will follow the unfolding method developed in \cite{Harvey:1995fq} for performing  this integral.
However we  generalize the discussion to integrands which contain modular 
forms transforming under $\Gamma_0(N)$ rather than $SL(2, \mathbb{Z})$. 
Let us first define the Fourier expansions of the expressions in the 
integrand. 
\begin{eqnarray} \label{defcrs} \nonumber
\frac{1}{2 \eta^{24} }E_4 
 \left[ \frac{1}{4} \alpha_{g'}^{(r,s)} E_6  - 
 \beta_{g'}^{(r,s)} (\tau) E_4 \right]\left[ 
 E_2 - \frac{3}{\pi \tau_2} \right]
&=& \sum_{l \in \mathbb{Z} / N }  \left( 
\tilde c^{(r, s)} ( l ) q^{l  } - 
\frac{3}{\pi \tau_2} c^{(r, s)}( l ) q^{l } \right), \\ 
&=& \tilde h{(r, s)} (\tau) - \frac{3}{\pi \tau_2}  f^{(r, s)} (\tau) .
\end{eqnarray}
Note that as expected, the twisted sectors admit fractional $q$ expansions. 
Substituting this expansion and the values  of $p_L^2$ and $p_R^2$ from 
(\ref{gamma22}) into (\ref{g1int}) we obtain 
\begin{eqnarray} \label{f1exp}
&& F_1(T, U)  = \sum_{r, s =0 }^{N-1} {\cal I }_{r, s}\;, \\ \nonumber
&&{\cal I}_{r,s} = \int_{\cal F}  \frac{d^2\tau}{\tau_2} 
\sum_{m_1, m_2, n_2\in \mathbb{Z}, n_1 \in \mathbb{Z} + \frac{r}{N} }
\exp\left[ 2\pi i \tau ( m_1 n_1 + m_2 n_2 ) \right] \times 
\\ \nonumber
&&\exp \left[ -\frac{\pi \tau_2}{T_2 U_2} \left| 
n_2  TU + n_1 T - U m_1 + m_2 \right|^2 \right]  e^{\frac{ 2\pi i s m_1}{N} }  
\left(  h^{(r, s)} (\tau) - \frac{3}{\pi \tau_2}  f^{(r, s)} (\tau) \right) .
\end{eqnarray}
The first step to do the  integral is to perform the Poisson re-summation over the 
momenta $m_1, m_2$ using the formula
\begin{equation}
\sum_{m \in \mathbb{Z}}  f(m) e^{2\pi i sm/N} 
= \sum_{k \in \mathbb Z+ \frac{s}{N} } \int_{-\infty}^\infty du f(u) \exp( 2\pi i k u) .
\end{equation}
Using this identity and performing the integral over the corresponding variables 
$u_1, u_2$, we obtain 
\begin{equation} \label{irs}
{\cal I}_{r, s} = \int_{\cal F}  \frac{d^2\tau}{\tau_2} T_2 
\sum_{n_2, k_2 \in \mathbb{Z}, n_1 \in \mathbb{Z} + \frac{r}{N}, 
k_1 \in \mathbb{Z} + \frac{s}{N}} 
\left( \tilde h{(r, s)} (\tau) - \frac{3}{\pi \tau_2}  f^{(r, s)} (\tau) \right) 
\exp\left[{\cal G} (\vec n, \vec k ) \right] ,
\end{equation}
where 
\begin{eqnarray}
{\cal G} ( \vec n, \vec k ) = - \frac{\pi T_2}{U_2 \tau_2} |{\cal A}|^2 - 2\pi i  T ( \rm{det} ) A , 
\\  \label{defA1}
A = \left( 
\begin{array}{cc}
n_1 & k_1 \\ n_2 & k_2 
\end{array}
\right) , \qquad
{\cal A} = ( 1, U ) A 
\left(
\begin{array}{c}
\tau \\ 1
\end{array}
\right) .
\end{eqnarray}
Therefore using (\ref{irs}) in (\ref{f1exp}) we can write the integral as 
\begin{eqnarray} \label{f1exp2}
F_1(T, U ) =   \int_{\cal F} ( \frac{d^2\tau}{\tau_2^2} \sum_{r, s=0}^{N-1} 
\sum_{n_2, k_2 \in \mathbb{Z}, \, n_1, k_1 \in \mathbb{Z} + \frac{r}{N}}
{\cal J} ( A, \tau) ,
\end{eqnarray}
where
\begin{eqnarray}\label{defJ}
{\cal J}(A, \tau)  = T_2 \exp\left(  - \frac{\pi T_2}{U_2}  |{\cal A}|^2 - 2\pi i T {\rm det} A 
\right) f^{(r, s)} ( \tau) \hat  E_2 (\tau) , \\ \nonumber
r=  Nn_1\;   {\rm mod } \;  N, \qquad s =  N k_1\;  {\rm mod} \; N.
\end{eqnarray}
Using (\ref{defA1}), we can think of ${\cal J}$ as a function of the matrix $A$. 
Then the sum over $r, s$ and $\vec n, \vec k$ in the right hand side of (\ref{f1exp2}) 
can be thought of as the sum over matrices of the form (\ref{defA1}) with 
$n_2, k_2 \in \mathbb{Z}$ and $n_1, k_1\in \frac{ \mathbb{Z}}{N}$. 
Thus (\ref{f1exp2}) can be written as 
\begin{equation}
I_1(T, U) =  \int_{\cal F}  \frac{d^2\tau}{\tau_2^2 } \sum_A {\cal J} ( A, \tau) .
\end{equation}
Now using the modular transformation given in (\ref{modtrans}) and the definition of 
${\cal J}$ in (\ref{defJ}), it can be seen that 
\begin{eqnarray}\label{jag1}
{\cal J} \left( A, \frac{a\tau + b}{c\tau + d} \right) = 
{\cal J} \left( A \left( 
\begin{array}{cc}
a & b \\
c& d
\end{array}
\right), \tau \right) .
\end{eqnarray}
This symmetry allows us to extend the integration over the fundamental domain to 
its images under $SL(2, \mathbb{Z})$ together with the restriction of the summation 
over $A$ to its inequivalent $SL(2, \mathbb{Z})$ orbits. 
Lets denote the sum over inequivalent 
$SL(2, \mathbb{Z})$ orbits as $\sum^{\prime}_A$, then $F_1$ becomes
\begin{eqnarray}\label{genusoneint}
{\cal I } = \sum_A^{\prime}  \int_{{\cal F}_A} 
\frac{d^2 \tau}{\tau_2^2} T_2 
\exp\left( -\frac{\pi T_2}{U_2 \tau_2} |{\cal A}|^2 - 2\pi i T{\rm det} A\right) 
f^{(r, s) }( \tau) \hat E_2 ( \tau) .
\end{eqnarray}
Now the label $r, s$ is to interpreted as $ N n_1\;   {\rm mod}\;  N $ and 
$N k_1 \; {\rm mod} \; N$ respectively.  The region of integration 
${\cal F}_A$ depends on the orbit represented by $A$.

From the analysis of \cite{Harvey:1995fq}  we see that there are three inequivalent orbits. 
These are as follows: the zero orbit
\begin{equation}
A= 0,
\end{equation}
the non-degenerate orbit
\begin{equation}
A = \left( \begin{array}{cc}
k& j \\ 0 & p 
\end{array}
\right) , \qquad   p \neq 0, p \in \mathbb{Z}, \qquad k, j \in \frac{1}{N} \mathbb{Z} , k > j
\end{equation}
and the degenerate orbit
\begin{equation}
A = \left(
\begin{array}{cc}
0 & j \\
0 & p
\end{array} 
\right), \qquad ( j, p) \neq ( 0, 0), \quad j \in \frac{1}{N} \mathbb{Z}, p\in \mathbb{Z}.
\end{equation}
The contribution from each of these orbits has been evaluated in the appendix \ref{detailg1}. 
Taking the  sum of these contributions the result for $F_1$ is given by 
\begin{eqnarray} \label{result}
&&F_1=\left.T_2 \frac{\pi}{6}E_2^2(q)f^{(0,0)}(q) \right|_{q^0}  \\ \nonumber
& +&\sum_{s=0}^{N-1}\left[\tilde{c}^{(0,s)}(0)
\left(-\log(T_2U_2)+\frac{2\zeta(2,s/N)}{\pi}U_2-\kappa\right)
-{c}^{(0,s)}(0)\frac{6\zeta(4,s/N)}{\pi^3} \frac{U_2^2}{T_2}+\frac{6}{\pi^2}{c}^{(0,s)}(0)\frac{\zeta(3)}{T_2U_2}\right]\\ \nn
&+&4{\rm Re}\left\{ \sum_{\begin{smallmatrix}
k\ge 0,l\geq-1\\(k,l)\ne (0,0)\\
k\in \mathbb{Z}+r/N
	\end{smallmatrix}}
	\left[ 
	\tilde{c}^{(r,s)}(kl)e^{2\pi i sl/N}{\rm Li}_1(e^{2\pi i (kT+lU)})-\frac{3}{\pi T_2U_2}c^{(r,s)}(kl)e^{2\pi i sl/N}{\tilde{\cal{P}}}(kT+lU) \right]  \right\},
\end{eqnarray}
where $\tilde{{\cal P}}$ is given by
\be
\tilde{\cal P}(x)={\rm Im}(x){\rm Li_2}(e^{2\pi i x})+\frac{1}{2\pi}{\rm Li_3}(e^{2\pi i x}).
\ee
From the appendix it will be clear that the above result holds for $T_2>U_2$ as
in the analysis of \cite{Harvey:1995fq}. 
Let us examine the result for $N=1$, the unorbifolded $K3\times T^2$ compactification. 
We obtain
For unorbifolded $K3$ this answer reduces to 
\begin{eqnarray} \label{resultk3}
F_1 
&=& -48\pi T_2 - 264 (-\log(T_2U_2) -\kappa) - 88\pi U_2 \\ \nn
&&16\pi \frac{U_2^2}{T_2}+\frac{6}{\pi^2}{c}(0)\frac{\zeta(3)}{T_2U_2}+\\ \nn
&&4{\rm Re}\sum_{\begin{smallmatrix}
	k\ge 0,l\geq-1\\(k,l)\ne (0,0)\\
	k\in \mathbb{Z}
	\end{smallmatrix}}\tilde{c}(kl){\rm Li}_1(e^{2\pi i (kT+lU)})-
	\frac{3}{\pi T_2U_2}c(kl){\cal{P}}(kT+lU).
\end{eqnarray}
Let us recall that for each of the orbifolds we can read out 
the coefficients $\tilde c^{(r, s)}$ and $c^{(r, s)}$ using their definition given  in 
(\ref{defcrs}) and the explicit expressions for the twisted elliptic genera given in 
the appendix \ref{frslist}.  
Let us make a few observations from the result for $F_1$. 
Note that the coefficients which determine the 
moduli dependence  of $F_1$ in the first two lines of (\ref{result}) depends on  low lying 
topological data of the new supersymmetric index.  This dependence 
does not involve exponentials in moduli $T, U$. 
The low lying coefficients of the new supersymmetric index  for the various
orbifold  compactifications corresponding to the Mathieu moonshine conjugacy classes
are listed in table . 

\begin{table}[H]
	\renewcommand{\arraystretch}{0.5}
	\begin{center}
		\vspace{0.5cm}
		\begin{tabular}{|c|c|c|}
			\hline
			& & \\
			Orbifold & $c^{(0,0)}(0)$ & $c^{(0,1)}(0)$ \\
			\hline
			\hline
			& &  \\
			2A & $-120$ & 136  \\
			3A & $-80$ & 109 \\
			5A & $-48$ & 77  \\
			7A & $-240/7$ & 829/14\\
			11A & $-240/11$ & 442/11\\
			23A & $-240/23$ & 473/23 \\
			\hline
		\end{tabular}
	\end{center}
\vspace{-0.2cm}
\caption{$c^{(0,0)}(0),\; c^{(0,1)}(0)$ are listed  for prime $N$.} \label{lowcoef1}
\renewcommand{\arraystretch}{0.5}
\end{table}
\begin{table}[H]
	\renewcommand{\arraystretch}{0.5}
	\begin{center}
		\vspace{0.5cm}
		\begin{tabular}{|c|c|c|c c|c|c|}
			\hline
			& & & & & & \\
			Orbifold & $c^{(0,0)}(0)$ & $c^{(0,1)}(0)$ & $s_1$ & $s_2$ & $c^{(0,s_1)}(0)$ & $c^{(0,s_2)}(0)$\\
			\hline
			\hline
			& & & & & & \\
			4B & $-60$ & 96 & 2 & \textemdash & 68 & \textemdash \\
			6A & $-40$ & 443/6 & 2 & 3 & 109/2 & 136/3 \\
			8A & $-30$ & 111/2 & 2 & 4 & 48 & 34 \\
			14A & $-120/7$ & 943/28 & 2 & 7 & 829/28 & 136/7 \\
			15A & $-16$ & 472/15 & 3 & 5 & 77/3 & 109/5  \\
			
			\hline
			\hline
			& & & & & &  \\
			2B & $-60$ & 124 & 2 & \textemdash& 196 & \textemdash\\
			3B & $-80/3$ & 166/3 & 3 & \textemdash & 407/6 & \textemdash \\
			& & & & & & \\
			\hline
			
		\end{tabular}
	\end{center}
	\vspace{-0.2cm}
	\caption{$c^{(0,0)}(0),\; c^{(0,1)}(0),\; c^{(0,s_1)}(0),\; c^{(0,s_2)}(0)$ are listed where $s_1$ and $s_2$ are divisors of the composite order $N$ with $s_1<s_2<N$.} \label{lowcoef2}
	\renewcommand{\arraystretch}{0.5}
\end{table}

One also observes that for an order $N$ orbifold $c^{(0,s)}(-1)=1/N$ and $\tilde{c}^{(0,s)}(0)$ can be written as 
\begin{equation}
\tilde{c}^{(0,s)}(0)={c}^{(0,s)}(0)-\frac{24}{N}.
\end{equation}
Using this topological data  we can evaluate the linear and quadratic dependences 
of the moduli $T$ and $U$ which result  from the contribution of the integral along the 
zero orbit and the degenerate orbit. 
 These coefficients for the various orbifolds are listed in table (\ref{lowcoef1}) and (\ref{lowcoef2}).
 Note that the coefficient of $\frac{\zeta(3)}{T_2U_2}$ is  determined  by the 
difference  $N_h-N_v$ or the Euler character. 
An  interesting observation  is the occurrence of the  low lying toplogical 
data weighted with the Hurwitz-zeta function $\zeta(2, s/N)$ as well as
$\zeta( 4, s/N)$ in the orbifolds.  This can be seen only for the orbifold compactifications. 
In fact if one is just given the  three coefficients of  $U_2$, and $U_2^2/T_2$ and $1/T_2 U_2$ in 
the gravitational threshold $F_1$ and the properties of the low lying coefficients
in , we can determine $c^{(0, s)}$ for all the orbifold compactifications 
corresponding to the conjugacy classes of Mathieu Moonshine considered  here. 
A summary of these coefficients, including the Euler number for all the orbifolds 
considered in this paper is given in  table \ref{chi}. 

The higher level coefficients  of the new supersymmetric index control the 
exponentially suppressed terms in the last line of (\ref{result}). Note that the 
exponential dependence of 
$T$ moduli carries the information of the twisted sectors.

\begin{table}[H]
	\renewcommand{\arraystretch}{0.5}
	\begin{center}
		\vspace{0.5cm}
		\begin{tabular}{|c|c|c|c|c|c|}
			\hline
			& & & & & \\
			Orbifold & $N_h-N_v$ & $\chi$ & $\sum_{s=0}^{N-1}\tilde{c}^{0,s}(0)(\frac{2\zeta(2,s/N)}{\pi})$  & $-\sum_{s=0}^{N-1}{c}^{0,s}(0)\frac{6\zeta(4,s/N)}{\pi^3} $ & $\frac{\pi}{6}E_2^2(q)F(q)|_{q^0}$\\ 
			& & & & & \\
			\hline
			\hline
			& & & & & \\
		1A & -240 & -480 & $-88\pi$ & $16\pi$ & $-48\pi$ \\
			& & & & & \\
			2A & 16 & 32 & $80\pi$ & $-128\pi$ & $-24\pi$\\
				& & & & & \\
			3A & 138 & 276 &$240\pi$ & $-576\pi$ & $-16\pi$ \\
				& & & & & \\
			5A & 260 & 520 & $560\pi$ & $-3200\pi$ & $-48\pi/5$\\
				& & & & & \\
			7A & 321 & 642 & $880\pi$ & $-9472\pi$ & $-48\pi/7$\\
				& & & & & \\
			11A & 380 & 760 &$1512\pi$ & $-39216\pi$ & $-48\pi/11$\\
				& & & & & \\
			23A & 442 & 884 & $3432\pi $& $-383664\pi$ & $-48\pi/23$\\
		\hline
			\hline
			& & & & & \\
			4B & 200 & 400 & $400\pi$& $-1600\pi$& $-12\pi$\\
				& & & & & \\
			6A & 262 & 524 & $720\pi$ & $-6240\pi$ & $-8\pi$ \\
				& & & & & \\
			8A & 322 & 644 & $1040\pi$ & $-15008\pi$& $-6\pi$\\
				& & & & & \\
			14A & 382 & 764 & $1992\pi$ &$-85584\pi$ & $-24\pi/7$ \\
				& & & & & \\
			15A & 382 & 764 & $2152\pi$ &$-105904\pi$ & $-16\pi/5$ \\
			
			\hline
			\hline
			& & & & &  \\
			2B & 384 & 768 & $640\pi$ & $-2176\pi$ & $-12\pi$ \\	
				& & & & & \\
			3B & 441 & 882 & $1428\pi$ & $-24264\pi$ & $-16\pi/3$\\
			& & & & & \\
			\hline
				
		\end{tabular}
	\end{center}
	\vspace{-0.2cm}
	\caption{$N_h-N_v,\; \chi$ \protect\footnotemark  and coefficients of $ U_2,\; U_2^2/T_2$ and $T_2,$ in the limit $T_2> U_2$ are shown.}
	\label{chi}
	\renewcommand{\arraystretch}{0.5}
\end{table}

\footnotetext{The Calabi Yau manifolds with $\chi$ values $-480,32,276,400,520,524,644,768$ are known to exist and they are listed in \cite{link}. }

\subsection{$g>1$}

The gravitational threshold  integral for $g>1$ is given by 
\begin{eqnarray}
F_g = \frac{1}{ \pi^2 ( 2 T_2 U_2) ^k}
\int_{{\cal F}} \frac{ d^2\tau}{\tau_2} 
\tau_2^{2k} 
  \frac{1}{\eta^{24} (\tau)} 
 \Gamma_{2,2}^{(r, s)}
 E_4 
 \left[ \frac{1}{4} \alpha_{g'}^{(r,s)} E_6  - 
 \beta_{g'}^{(r,s)} (\tau) E_4 \right] ( p_R^{(r,s)} )^{2k} {\cal P}_{2k +2} .\nonumber \\
\end{eqnarray}
where $ k  =g -1>0$. 
Note that here the summation over $(r, s)$ is implied. 
The steps to perform the integral are similar to the case when $g=1$, 
however one needs to keep track of the extra insertions of the momentum 
$p_R^{(r,s)}$.  
First we perform the Poisson re-summation over the variables 
$(m_1, m_2)$ to obtain
\begin{eqnarray}\label{fgexp2}
F_g(T, U ) = \sum_{r, s=0}^{N-1}  
 \sum_{n_2, k_2 \in \mathbb{Z}, 
n_1, k_1 \in \mathbb{Z} + \frac{r}{N} }
 ( \frac{T_2}{2U_2})^k   \int_{\cal F} \frac{d^2\tau}{\tau_2^2} 
\tilde{{\cal J} } ( A, \tau) , \\ \label{deftJ}
\tilde{ {\cal J} }  ( A, \tau) = T_2 {\cal A}^{2k}\exp
\left(  - \frac{\pi T_2}{U_2}  |{\cal A}|^2 - 2\pi i T {\rm det} A 
\right) f^{(r, s)} ( \tau)  {\cal P}_{2k + 2}( \tau)  , \\ \nonumber
r=  Nn_1\;   {\rm mod } \;  N, \qquad s =  N k_1\;  {\rm mod} \; N.
\end{eqnarray}
Now similar to the $g=1$ case we can think of the sum over 
$r, s$ and $\vec n, \vec k$  as sum over the matrices 
of the form in 
(\ref{defA1}) with 
$n_2, k_2 \in \mathbb{Z}$ and $n_1, k_1\in \frac{ \mathbb{Z}}{N}$. 
Thus (\ref{fgexp2}) can be written as 
\begin{equation}
I_g(T, U) =  \int_{\cal F}  \frac{d^2\tau}{\tau_2^2 } \sum_A \tilde{ {\cal J}} ( A, \tau) .
\end{equation}
Now using the modular transformation given in (\ref{modtrans}) and the definition of 
$\tilde{\cal J}$ in (\ref{deftJ}), it can be seen that 
\begin{eqnarray}\label{ja}
\tilde{{\cal J}} \left( A, \frac{a\tau + b}{c\tau + d} \right) = 
\tilde{{\cal J}} \left( A \left( 
\begin{array}{cc}
a & b \\
c& d
\end{array}
\right), \tau \right) .
\end{eqnarray}
This symmetry allows us to extend the integration over the fundamental domain to 
its images under $SL(2, \mathbb{Z})$ together with the restriction of the summation 
over $A$ to its inequivalent $SL(2, \mathbb{Z})$ orbits. 
Let us again denote the sum over inequivalent 
$SL(2, \mathbb{Z})$ orbits as $\sum^{\prime}_A$, then $F_g$ becomes
\begin{eqnarray} \nonumber
F_g= \sum_A^{\prime}  \int_{{\cal F}_A} ( \frac{T_2}{2U_2})^k 
\frac{d^2 \tau}{\tau_2^2} T_2 {\cal A}^{2k}
\exp\left( -\frac{\pi T_2}{U_2 \tau_2} |{\cal A}|^2 - 2\pi i T{\rm det} A\right) 
f^{(r, s) }( \tau) {\cal P}_{2k + 2}( \tau). \\ \label{sumorb}
\end{eqnarray}
Now the label $r, s$ is to interpreted as $ N n_1\;   {\rm mod}\;  N $ and 
$N k_1 \; {\rm mod} \; N$ respectively.  The region of integration 
${\cal F}_A$ depends on the orbit represented by $A$. 

We can now look at  the contribution of the three inequivalent orbits. 
First note that the  contribution of the zero orbit vanishes due to the 
presence of ${\cal A}^{2k}$ in the integrand. 
Thus we are left with the non-degenerate orbit 
which is characterized by the set of matrices 
\begin{equation}
A = \left( \begin{array}{cc}
n_1 & j \\ 0 & p 
\end{array}
\right) , \qquad    p \in \mathbb{Z}, \qquad n_1, j \in \frac{1}{N} \mathbb{Z} ,\; n_1 > j
\end{equation}
and the degenerate orbit
\begin{equation}
A = \left(
\begin{array}{cc}
0 & j \\
0 & 0
\end{array} 
\right),  \qquad  j \in \frac{1}{N} \mathbb{Z},   j \neq 0 .
\end{equation}
Note that here we have included the $p=0$ case also  in the non-degenerate orbit for convenience. 
This is because the $p=0$ can be treated uniformly together with the the $p\neq 0$ situation in 
the non-degenerate orbit.

The detailed evaluation of the integral in the two orbits is carried out in the appendix \ref{detailg2}. 
This integral was done in \cite{Marino:1998pg} for the case of $K3\times T^2$ compactification
and in \cite{Klemm:2005pd} for the FHSV compactification. The latter situation involves 
only modular forms under $\Gamma_0(2)$ in the integrand. 
Here we have generalized the evaluation of the integral for integrands containing 
modular forms transforming under $\Gamma_0(N)$. 
Further more as it will be clear in the appendix \ref{detailg2}, we do not directly apply the 
reduction theorem of Borcherds \cite{Borcherds:1996uda} 
as done in the earlier works to perform the integral. 
The application of the reduction theorem is rather intricate and it 
easier to proceed 
straightforwardly from (\ref{sumorb}) and carry out the steps involved in the integration.

To write  the result in a convenient form, let us define 
the following two dimensional vectors and the inner products. 
\begin{eqnarray}\label{notation}
&& m  = ( n_1, n_2 ) , \quad n_1 \in \mathbb{Z} + \frac{r}{N}, \; 
n_2 \in \mathbb{Z}, \qquad 
 \frac{m^2}{2} = n_1n_2,   \nonumber \\ 
 & &  y = ( T, U ) , \\ \nonumber
& & m \cdot   y = n_1T_1 + n_2 U_2  + i (n_1 T_1 + n_2 T_2 ) , 
\\ \nonumber
&& 
 m \hat{\cdot} y = n_1 T_1 + n_2 U_2 - i (n_1 T_1 + n_2T_2 ) .
\end{eqnarray}
We also define the Fourier coefficients involved in the 
expansion of the modular forms as
\begin{equation}\label{defcg}
f^{(r, s)} ( \tau) {\cal P}_{2k +2} (\tau)  = 
\sum_{l \in \frac{\mathbb{Z}}{N} , t = 0 }^{t = g} c_{g-1}^{(r,s)} ( l, t) \tau_2^{-t} q^l .
\end{equation}
Then,  from the evaluation in  appendix \ref{detailg2}, the result for the contribution from 
the  non-degenerate orbit   is given by 
\bea \label{fresultfg}
F_{g>1}^{ \rm{nondeg} } &=& \frac{(-1)^{g-1}}{2^{2(g-1)}\pi^2}\sum_{r,s=0}^{N-1}\sum_{m\ne 0}\sum_{t=0}^g \sum_{h=0}^{2g-2}\sum_{j=0}^{g-1-h/2}\sum_{a=0}^{\hat s}
 e^{-2\pi i s n_2/N}c^{(r,s)}(m^2/2,t) \times \nonumber \\ \nn
&& \frac{(2g-2)!}{j!h!(2g-h-2j-2)!} \frac{(-1)^{j+h}}{(4\pi)^{j+a}}\frac{(\hat s+a)!}{(\hat s-a)!a!}({\rm sgn}({\rm{Im} }(m\cdot y)))^h \times \\ \nn
&&\frac{({\rm{Im}}(m\hat \cdot y))^{t-j-a}}{(T_2U_2)^t} {\rm Li}_{3+a+j+t-2g}(e^{2\pi i m\hat \cdot y})\\ \nn
&+& \sum_{s=0}^{N-1} \,\frac{c^{(0,s)}(0,g-1)}{2^{4g-5}\, \pi^{g+1}}\frac{1}{(T_2U_2)^{g-1}} \sum_{\tilde s=0}^{g-1}(-1)^{\tilde s} \frac{(2g-2)!}{\tilde{s}!(g-1-\tilde s)!}\psi (1/2+\tilde {s})\\ \nn
&+& \sum_{s=0}^{N-1}\sum_{t=0}^{g-2}\,\frac{c^{(0,s)}(0,t)}{\pi^{t+5/2}} \, \frac{\zeta(3+2(t-g))}{(T_2U_2)^t}
\times \\ 
&& \sum_{\tilde s=0}^{g-1}(-1)^{\tilde s} 2^{2(\tilde s-2g+2)}\frac{(2g-2)!}{(2\tilde s)!(g-1-\tilde s)!}\Gamma(3/2+\tilde s+t-g). 
\eea
Here 
\begin{equation}\label{hatsdef}
\hat s + 1/2 = |\nu|, \qquad \qquad |\nu| = 2k - h - j - t - 1/2, \qquad k = g-1
\end{equation}
and  $m \neq 0$ refers to  any  of the following cases,
\bea \label{limits}
&&n_1>0, n_2>0; \qquad n_1<0, n_2<0;\\ \nn
&&  n_1=0, n_2>0\; {\rm or} \; n_2<0; \qquad  n_2=0, n_1>0 \; {\rm or} \; n_1<0 ;\\ \nn
&& n_1=\frac{r}{N} , n_2 <0 \; \hbox{or} \;
n_1=-\frac{r}{N}, n_2>0 \; \hbox{with}\;\; r |n_2| \leq N, r>0.
\eea
In (\ref{fresultfg}), the index $r$ in $c^{(r, s)}$ is related to  $n_1$ by 
$r =N n_1$ \;   {\rm mod}\, N and $n_2\in \mathbb{Z}$ .
The result for the non-degenerate orbit is valid for $T_2, U_2 >0$ \footnote{
Note that the result for the integral also contains the complex  conjugate $\bar{F_g}$, which 
we have  suppressed for simplicity. }. 
Performing the integral for the degenerate orbit we obtain 
for $T_2<U_2$ 
\be\label{fresultdeg}
F_{g>1}^{{\rm deg}}=\frac{1}{T_2^{2g-3}}\sum_{s=0}^{N-1}\sum_{t=0}^g c^{(0,s)}(0,t)\frac{t!}{2^{2g-2}\pi ^{t+3}}\left(\frac{T_2}{U_2}\right)^t\zeta(2(2+t-g), \frac{s}{N}).
\ee
The   result for the integral in the degenerate orbit for $T_2>U_2$ is obtained by interchanging
$T_2 \leftrightarrow U_2$ in (\ref{fresultdeg}).

\subsection{Extracting the holomorphic $F_g$}

We now follow \cite{Marino:1998pg} and extract the part of  the one loop gravitational
coupling given in (\ref{fresultfg}) and (\ref{fresultdeg}) which does not mix the holomorphic and anti-holomorphic
parts in the moduli $T, U$. 
This part of the gravitational coupling can be written in terms of Gopakumar-Vafa invariants
and essentially counts holomorphic curves in the target space 
\cite{Gopakumar:1998ii,Gopakumar:1998jq}. 
Note the non-degenerate contribution given in (\ref{fresultfg}) along includes its
complex conjugate. 
It is clear that the degenerate orbit does indeed mixes the holomorphic
and anti-holomorphic parts of the moduli, and therefore cannot be written as a purely holomorphic term. 
Examining the non-degenerate orbit in (\ref{fresultfg}), we see that the 
holomorphic term can only arise from $j=t=a=0$.  Then only the function ${\rm Li}_{3-2g}$
contributes. 

Now if $n_1\geq 0, n_2 \geq 0 $  then  one obtains the sum 
\be
\sum_{h=0}^{2g-2}\frac{(2g-2)!}{h!(2g-2-h!)}(-1)^h ({\rm sgn} ({\rm Im}(m\cdot y)))^h =0 .
\ee
Thus for these lattice points there is no holomorphic contribution \footnote{ 
Similarly for $n_1 \leq 0, n_2 \leq 0$, then there is no anti-holomorphic contribution 
from the complex conjugate of $F_g$. }.
There is a non-zero contribution from $n_1<0, n_2 <0$ from the $F_g$ as given 
in  (\ref{fresultfg}). Once can then change the sum over $n_1, n_2$ by flipping their sign
and the result is that one obtains a purely anti-holomorphic term \footnote{
A similar observation can be made for the complex conjugate of $F_g$. }. 
Thus the contribution from these lattice points can be written as a purely anti-holomorphic
term.  The lattice points $n_1 = 0, n_2 \leq 0$ as well as 
$n_1<0, n_2 =0$ also contribute to terms that do not have any mixing. 

The remaining contribution which does not have mixing arises from the lattice points
\bea \label{lim2sp}
 &&  n_1=\frac{r}{N} , n_2<0, r>0\;\qquad  {\rm if}\;\;\;
 \frac{r}{N} T_2< n_2 U_2;  \\ \nonumber
 & &  n_1=-\frac{r}{N} , n_2>0 , r>0 \;\qquad  {\rm if}\;\;\;\frac{r}{N}T_2> n_2 U_2;\\ \nonumber
 & &{\rm with} \;\;  r |n_2| \leq N .
\eea
Note that these lattice points also  include $(-1, 1)$ and $( 1, -1)$.
The contribution from    points $(n_1, n_2 ) =  ( r/N,  n_2); n_2 <0, r>0$ is non-vanishing only 
 for $r/NT_2<  n_2U_2$  and  result in ${\rm Li}_{3-2g} ( q) $ where
$q = \exp [ 2\pi i ( r/N T_1 - n_2 U_1)  +2\pi ( r/N T_2 - n_2 U_2) ]$. 
Similarly the contribution from the points  $(n_1, n_2 ) =  ( r/N,  n_2); n_2 >0, r>0$
is non-vanishing only for   $ r/NT_2<  n_2U_2$  and results in ${\rm Li}_{3-2g} ( q^{-1}) $. 
These two contributions can be combined using the relation
\be
{\rm Li}_{a}(z)=(-1)^{|a|+1}{\rm Li}_{a}(z^{-1}).
\ee
which is valid for $a<0$ 
into a  single holomorphic function for the lattice points $(n_1, n_2 ) =  ( r/N,  n_2); n_2 >0, r>0$
 and valid for any $T_2, U_2$.
Finally the last term in (\ref{fresultfg}) contributes at $t=0$ as a constant.
We can perform the sum over  $\tilde s$ using the identity 
\begin{eqnarray}
\sum _{\tilde s=0}^{g-1} \frac{(-1)^{\tilde s} (2 g-2)! 2^{2 (-2 g+\tilde s+2)} 
}{(2 s)! (g-\tilde s-1)!} \Gamma (-g+\tilde s+\frac{3}{2})
=\frac{1}{2} \sqrt{\pi }(-1)^{g-1}.
\end{eqnarray}

Using all these arguments we obtain the following expression for the 
purely holomorphic contribution to the gravitational couplings. 
\bea\label{fhol}\nonumber
\bar{F}_g^{{\rm hol}}=\frac{(-1)^{g-1}}{\pi^2} \sum_{s=0}^{N-1} \left( \sum_{m>0}e^{-2\pi i n_2s/N}c^{(r,s)}_{g-1}(m^2/2,0){\rm Li}_{3-2g}(e^{2\pi im\cdot y})+\frac{1}{2}c_{g-1}^{(0,s)}(0,0)\zeta(3-2g)
\right). \\
\eea
In the sum over lattice points, $m>0$ refers to  the following lattice points 
\begin{eqnarray}
n_1, n_2 \geq 0, \quad\hbox{but } ( n_1, n_2 ) \neq ( 0, 0 ), \\ \nonumber
  ( r/N, - n_2) , \qquad \hbox{with}\;\;  n_2>0 \;\;\hbox{and}\;\; rn_2 \leq N .
\end{eqnarray}

Though the  holomorphic contribution was obtained by examining the 
situation for $g>1$, the result can be extrapolated for $g=1$ as well
as $g=0$. For $g=1$ we have evaluated the gravitational coupling
explicitly in (\ref{result}). By examining this result we see that 
except for  the constant, the holomorphic part in (\ref{fhol})  agrees with 
(\ref{result}) up to a normalization for $g=1$.  For this, note that from the definition 
of $\tilde c^{(r,s)}$ in (\ref{defcrs}) and $c_{0}^{(r,s)} (m^2/2, 0) $ in (\ref{defcg}), 
they are related to each other by a normalization. 
The constant in (\ref{fhol}) for $g=1$ requires a regularization to agree with the constant
in (\ref{result}) proportional to  $\kappa\sum_{s=0}^{N-1} \tilde c^{(0,s)}(0)$.

\section{Gopakumar-Vafa invariants and  conifold singularities}\label{GVCS}

The $E_8\times E_8$ heterotic string compactified on $K3\times T^2$ is dual to 
type II A compactified on a  known  Calabi-Yau  geometry \cite{Kachru:1995wm}  by 
${\cal N}=2$ string duality. 
The compactifications which are the focus of study in this paper are 
presumably dual  to appropriate freely acting  quotients of this space
which needs to be discovered. 
However since we have evaluated the gravitational couplings $\bar F_g^{\rm hol} $  from 
the heterotic side, we do have some topological information of the dual Calabi-Yau.  
To obtain this we cast the $\bar F_g^{\rm hol} $  in to a form from which 
we can read out the Gopakumar-Vafa invariants associated with the Calabi-Yau. 
The generating functional for the $F_g$'s can be written as
\begin{eqnarray} \label{genfGV}
F^{{\rm GV}} (g_s, y) &=& \sum_{g=0}^\infty F^{\rm GV}_{g} (y ) g_s^{ 2g -2}, \\ 
\nonumber
& =& 
 \sum_{g=0}^\infty \sum_{ m>0} \sum_{d=1}^\infty
n_{m}^g \frac{1}{d} \left( 2\sin \frac{dg_s}{2} \right)^{2g-2} 
e^{2\pi i d (   m\cdot y  ) }.
\end{eqnarray}
Here we have focused on only the $y = (T, U) $ moduli. 
In general 
they will include all the K\"{a}hler parameters of the 
dual Calabi-Yau.  The quantities $n_{m}^g$ are called
Gopakumar-Vafa invariants and they are integer numbers 
which encodes the number of 
BPS states which arise in the compactification of the type IIA  side of the 
theory. For a nice review see \cite{Marino:2002wa} \footnote{We thank Rajesh Gopakumar
for bringing this review to our attention.}.   

It is not apriori obvious that the 
compactifications on orbifolds  of $K3 \times T^2$ associated with 
moonshine symmetry 
 we have considered here 
will admit integer Gopakumar-Vafa invariants. 
In this section we will observe that, this is indeed the 
case for low values of $m = ( n_1, n_2) $ for genus 0, 1 and 2. 
We will then study the points  in the moduli space at which 
the topological amplitude $\bar F_g^{\rm hol} $  becomes singular.
On the heterotic side, these are points at which the certain hyper multiplets
become massless. While on the type II A side,  
these points are associated with the conifold singularity. 
We will determine the points in moduli space at which these  singularities occur. 
We observe that the states which become massless at these points in 
moduli space occur  only in the twisted sector for these compactifications
and the strength is determined by the genus zero Gopakumar-Vafa invariant for these 
sectors. 

\subsection{Low lying coefficients of $\bar F_g^{\rm hol}$}

Before we proceed, it is useful to determine the properties of  certain low lying   coefficients 
$c_{g-1}^{(r,s)}$ that occur in $\bar F_g^{\rm{hol}}$. 
  For this we obtain  equations relating  the constant in (\ref{fhol})
  $ \sum_{s=0}^{N-1} 
  c_{g-1}^{(0,s)} ( 0, 0) $ and
  the Fourier coefficients  
  $ \sum_{s=0}^{N-1} c^{(0, s) }(0 )$ in the expansion (\ref{defcrsfrs}). 
  Note that as we are in the untwisted sector the Fourier expansion is in integer powers of 
  $q$.  To extract this coefficient we need to examine the terms independent of 
  $\tau_2$ in the expansion (\ref{defcg}). 
  Therefore we can  ignore the $\tau_2$ dependence in the 
  generating function (\ref{thetexp}). The generating function  reduces 
  as
  \begin{eqnarray}\label{seqrel}
  \sum_{k=0}^\infty \tilde \lambda^{2k} {\cal P}_{2k} (  G_2, \ldots, G_{2k})  &=& 
-\left(\frac{2\pi \eta^3 \tilde \lambda }{\theta_1(\tilde\lambda ,\tau)}\right)^2, \\ \nonumber
&=&- \left(\frac{\pi \tilde\lambda}{\sin \pi \tilde\lambda}\right)^2+8\pi^2\tilde\lambda^2 q + \cdots , \\ \nonumber
&=& \sum_{g=0}^{\infty}\frac{(2g-1)}{(2g)!}(-1)^g(2\pi \tilde\lambda)^{2g}B_{2g} +8\pi^2\tilde\lambda^2 q + \cdots, 
\end{eqnarray}
where $B_{2g}$ are Bernoulli numbers.  These can be related to  the $\zeta$ function using
\begin{equation}
B_{2g}=(-1)^{g+1}\zeta(2g)\frac{2 (2g)!}{(2\pi)^{2g}}.
\end{equation}
Using (\ref{seqrel}) we see that 
\begin{equation}\label{e2}
{\cal P}_0 = -1, \qquad
\left. {\cal P}_{2}( G_2)\right|_q = 8\pi^2, \qquad
\left. {\cal P}_{2g}( G_2,...,G_{2g}) \right|_q = 0, \; \hbox{for}\; g>1 .
\end{equation}
Now in the untwisted sector, we have the expansion
\begin{equation}\label{e1}
f^{(0, s)} ( q) = c^{(0, s) }( -1) q^{-1} + c^{(0, s) }(0)  +  c^{(0, s) }(1) q + \cdots
\end{equation}
Now using the definition (\ref{defcg})  and the equations
(\ref{seqrel}), (\ref{e2})  and (\ref{e1}) we obtain
\begin{eqnarray} \label{relcsco}
\sum_{s=0}^{N-1} 
  c_{g-1}^{(0,s)} ( 0, 0) &=& 
  \sum_{s=0}^{N-1}c^{(0,s)}(0)\frac{(2g-1)}{(2g)!}(-1)^g (2\pi)^{2g}B_{2g},\\ \nn
&=&\frac{\chi(X)}{2} \frac{(2g-1)}{(2g)!}(-1)^g (2\pi)^{2g}B_{2g}, \qquad g>1 .
\end{eqnarray}
In the second line we have used the definition of the Euler character. 
Now for  $g=1$ we obtain
\begin{eqnarray}
\sum_{s=0}^{N-1} 
  c_{0}^{(0,s)} ( 0, 0) &=& - 2\pi^2 B_2 \sum_{s=0}^{N-1}c^{(0,s)}(0)  +  
  8\pi^2\sum_{s=0}^{N-1} c^{(0,s)}(-1) , \\ \nonumber
  &=& - \frac{\pi^2}{12}  \chi + 8\pi^2,
  \end{eqnarray}
  where we have used $c^{(0,s)}(-1) = \frac{1}{N}$, which can be shown to be 
  true for all the orbifolds
  And finally for $g=0$ we obtain
  \begin{equation} \label{g0rel}
  \sum_{s=0}^{N-1} 
  c_{-1}^{(0,s)} ( 0, 0) = - \sum_{s=0}^{N-1}c^{(0,s)}(0) = - \frac{\chi}{2}.
  \end{equation}
 Still being within the untwisted sector and 
  using the same methods we obtain the relation
    \begin{eqnarray}\label{untwissi}
\sum_{s=0}^{N-1} e^{ \frac{ 2\pi i s}{N}  }
  c_{g-1}^{(0,s)} ( -1, 0) &=& 
  \sum_{s=0}^{N-1} e^{ \frac{ 2\pi i s}{N}  } c^{(0,s)}(-1)
  \frac{(2g-1)}{(2g)!}(-1)^g (2\pi)^{2g}B_{2g},  \\ \nonumber
  &=&  0, \;\;\hbox{for all} \;\; g .
\end{eqnarray}
To obtain the last line we have used   the fact hat 
$c^{(0, s)}(-1) = \frac{1}{N}$ for all the orbifolds. 

  Let us now proceed similarly and relate the coefficients of the low lying 
  twisted sector. 
  From examining the expansions of $f^{(r, s)}$ defined in 
  (\ref{defcrsfrs}), we see  from the explicit evaluation of the $q$-expansions that the
  non-zero 
  leading terms are  given by 

  \begin{equation}
  \sum_{s=0}^{N-1} e^{ 2\pi i \frac{n_2 s}{N} }  f^{(r, s)} (q) = 
  \sum_{s=0}^{N-1} e^{ 2\pi i \frac{n_2 s}{N} }c^{(r,s)} ( -\frac{rn_2}{N} ) q^{- \frac{rn_2}{N} }  + 
  O( q^{1- \frac{rn_2}{N} } ) .
  \end{equation}
  Here $r, n_2>0$ and $rn_2\leq N$. Therefore using the same methods we obtain the relation
  \begin{equation}\label{coni1}
  \sum_{s=0}^{N-1} e^{ 2\pi i \frac{n_2 s}{N} }  c_{g-1}^{(r, s) } ( -\frac{rn_2}{N}, 0 ) 
  =   \sum_{s=0}^{N-1} e^{2\pi i \frac{n_2 s}{N} } c^{(r,s)} ( -\frac{rn_2}{N}) 
  \frac{(2g-1)}{(2g)!}(-1)^g (2\pi)^{2g}B_{2g}.
  \end{equation}
  
  Let us define the following quantity 
  \begin{equation} \label{g0coni}
  n^0_{(\frac{r}{N}, -n_2)} \equiv
   -2 \sum_{s=0}^{N-1} e^{ 2\pi i \frac{n_2 s}{N} } c^{(r,s)} ( -\frac{rn_2}{N}) .
  \end{equation}
  Evaluating this for all the orbifold compactification of the heterotic string 
  considered in this paper and in each of  the twisted sectors, we see that 
 $ n^0_{(\frac{r}{N}, -n_2)}$ is an integer.  We will see that this quantity 
 will turn out to be the genus zero Gopakumar-Vafa  invariant  at the point 
 $m = (\frac{r}{N}, -n_2)$.  We list them in appendix \ref{GVZ} .

 \subsection{Integrality of the Gopakumar-Vafa invariants}
 
 To begin we first  write down the Gopakumar-Vafa  form of the genus $g$  topological 
amplitude from (\ref{genfGV}). 
For $g>1$, this is given by 
\bea \label{fgv}
 F^{{\rm GV}}_g&=&\frac{(-1)^g |B_{2g}B_{2g-2}|\chi(X)}{4g (2g-2)(2g-2)!}\\ \nn
 &+& \sum_{\beta}\left[\frac{|B_{2g}|n_m^0}{2g (2g-2)!}+\frac{2(-1)^g n_m^2}{(2g-2)!}\pm...-\frac{g-2}{12}n_{m}^{g-1}+n_m^g\right]{\rm Li}_{3-2g}(e^{2\pi i m\cdot y}).
 \eea
 Here we have included the constant term which is  the contribution to the topological 
 amplitude due to holomorphic maps from genus $g$ surface to a single point. 
 For $g=0$ we have we obtain  
 \begin{equation} \label{fg0gv}
 F_0^{\rm GV} = \zeta(3) \frac{\chi(X)}{2} + \sum_{m>0} n_{m}^0 {\rm Li}_3( e^{2\pi i m\cdot y}),
 \end{equation}
 where we can included the contribution due to the Euler characteristic of the Calabi-Yau target.
 Finally for $g=1$ we have
 \begin{equation} \label{f1gv}
 F_1^{\rm GV} = \sum_{m>0} \left( \frac{1}{12} n_m^0 + n_m^1 \right) {\rm Li}_{1}  
 ( e^{2\pi i m\cdot y}).
 \end{equation}
 Here we have kept only the instanton contribution to the topological amplitude. 
 The reason we have not kept the constant  is that we  need to regulate
 the constant term  in (\ref{fhol}) to extract this contribution for $g=1$ to compare 
 with the expected results from the type II Calabi-Yau side. 
 Note that on examining the constant  term in the  calculation 
 done to obtain $F_1$ in (\ref{result})  the  regulator  we used  essentially involves 
 subtracting the divergence in $\zeta(1)$ to obtain the Euler-Mascheroni constant.

 We can now compare the constant term in (\ref{fhol}) and (\ref{fgv}) to fix the overall 
  normalization relating   $\bar F_g^{\rm hol}$ and $F^{{\rm GV}}_g$ for $g>1$ 
  For this we need to use  the relation (\ref{relcsco}) and identities relating $\zeta(3-2g)$
  to Bernoulli numbers. 
  Requiring the constant term to agree results in the following relation between the 
  topological amplitudes
  \begin{equation}\label{relfgfhol}
  F_g^{{\rm GV}} =    \frac{ (-1)^{g+1}}{2 ( 2\pi )^{2g-2} } \bar F_g^{\rm hol}.
  \end{equation}
  Though this relation was obtained  for $g>1$ we will  also extrapolate it for $g= 1, 0 $.
  Using the relation in (\ref{relfgfhol}) we can relate the Gopakumar-Vafa 
  invariants $n_m^g$ to the coefficients $c_{g-1}^{(r, s) }( m^2/2, 0 ) $ . 
  We will now perform this explicitly for $g=0, 1, 2, 3$ and will observe that the 
  resulting Gopakumar-Vafa invariants are integral.

  \vspace{.5cm}
  \noindent 
  $g=0$
  \vspace{.5cm}
  \\
  For $g=0$ we have the relation
  \begin{eqnarray}
  F_0^{\rm GV} = -2\pi^2 \bar F_0^{\rm hol}.
  \end{eqnarray}
   We see that  using (\ref{g0rel}), the constant term in 
  (\ref{fg0gv}) and (\ref{fhol})  agrees. 
   Furthermore we can identify
  \begin{eqnarray}\label{g0gvinv}
  n^0_{(n_1, n_2)} &=&  2 \sum_{s=0}^{N-1} e^{-\frac{2\pi i n_2 s }{N} } c_{-1}^{(r, s) }(m^2/2, 0),
   \qquad r   =  n_2 N  \; {\rm mod} \; N
  \\ \nonumber &=&  -2 \sum_{s=0}^{N-1} e^{-\frac{2\pi i n_2 s }{N} } c^{(r,s)} ( n_1n_2 ) .
  \end{eqnarray}
  To obtain the last line we have used (\ref{defcg}) and the fact that ${\cal P}_0=-1$.
  Note that a special case of this identification is 
  that given in (\ref{g0coni}). 
  We explicitly evaluate the genus zero  low lying Gopakumar-Vafa invariants for 
  for all the $\mathbb{Z}_N$ orbifolds of  $K3\times T^2$    considered in this paper
  and observe that these are integers. 
  A listing of these are provided in tables. 
  
  Now we will demonstrate that if the $n^0_m$ invariants are integers, then 
  it is guaranteed that Gopakumar-Vafa invariants for $g=1, 2, 3$ are integers. 
  This analysis  most likely extends to all genera.      
  
  \vspace{.5cm}
  \noindent 
  $g=1$
  \vspace{.5cm}
  \\
For $g=1$ the relation between the Gopakumar-Vafa form of the 
topological amplitude and that obtain from the threshold integral is 
given by 
\begin{equation}
F_1^{\rm GV} = \frac{1}{2}  \bar F_1^{\rm hol}.
\end{equation}
Comparing the instanton contributions of $(\ref{fhol})$ and $(\ref{f1gv})$, we see that 
we can identify 
\begin{equation}
n^1_{(n_1, n_2)} = \frac{1}{2\pi^2} \sum_{s=0}^N e^{-\frac{2\pi i n_2 s}{N}} c^{(r,s)}_0 ( m^2/2, 0 ) 
- \frac{1}{12} n^0_{(n_1, n_2) }.
\end{equation}
Now using (\ref{g0gvinv}) and the definition (\ref{defcg}) for $g=0, 1$  together with 
the relation
\begin{equation}
{\cal P}_2(G_2) = - 2 \frac{\pi^2}{6} E_2,
\end{equation}
we find that 
that the genus one Gopakumar-Vafa invariants are  guaranteed to be integral if the 
following are satisfied.  
\begin{enumerate}
\item
The 
Fourier coefficients of the following quasi-modular form are 
integral.
\begin{equation}
-\frac{1}{6} ( {E_2}(\tau)  - 1)  = 4 \sum_{l =1}^\infty \sigma_1(l ) q^ l .
\end{equation}
Here $\sigma_1(l)$ is the sum of divisors of $l$. 
\item
The  genus zero Gopakumar-Vafa invariants $n^0_m$ are integers.
\end{enumerate}
The condition (1) is  assured by the definition of $E_2(\tau)$. 
We observe that the  condition (2) holds true for all the orbifolds we consider. 
Therefore the genus one Gopakumar-Vafa invariants are assured to be 
integers.

  \vspace{.5cm}
  \noindent 
  $g=2$
  \vspace{.5cm}
  \\
  For this case the relation between the amplitudes becomes
  \begin{equation}
  F_2^{\rm GV} =- \frac{1}{8\pi^2} \bar F^{\rm hol}_2.
  \end{equation}
  Comparing the instanton contributions it is easy to see that 
  \begin{equation}
  n^2_{(n_1, n_2) } =   \frac{1}{8\pi^4} \sum_{s=0}^{N-1} e^{- \frac{2\pi i n_2}{N}} 
  c_1^{(r, s)}( m^2/2, 0) - \frac{|B_4|}{8} n_{(n_1, n_2)}^0.
  \end{equation}
  Using the definitions we get 
  \begin{equation}
  {\cal P}_4  = - \pi^4 \left( \frac{1}{18} E_2^2 + \frac{1}{90} E_4 \right).
  \end{equation}
  Then from (\ref{g0gvinv}) and the definition (\ref{defcg}) for $g=0, 2$  we can see that 
  $n_m^2$ are integers provided the genus zero invariants $n_m^0$ are integral
 and the following quasi-modular form admits integral Fourier coefficients
 \begin{equation}
 \frac{1}{720} ( 6 - 5 E_2^2 - E_4)  = \sum_{l} n_l q^l .
 \end{equation}
 For $n_l \in \mathbb{Z}$, we must have that 
   the Fourier coefficients of $( 6 - 5 E_2^2 - E_4)$ be divisible by $720$. 
   This can be verified easily from the definition of the Eisenstein series
   \begin{equation}
    6 - 5 E_2^2 - E_4 = 
 240 \sum_k (\sigma_1(k)-\sigma_3(k))q^k+ 2880 \sum_{k_1,k_2} \sigma_1(k_1) \sigma_1(k_2)  q^{k_1+k_2}.
   \end{equation}
  The last term is obviously divisible by $720$. The first term can be written as
  \begin{equation}
  240 \sum_k (\sigma_1(k)-\sigma_3(k)) = 240\sum_{d|k} (d-1)(d)(d +1).
  \end{equation}
  Now  $(d-1)(d)(d +1) $ is always a multiple of $6$ and therefore the first term is 
  also divisible by $720$. 
  Thus the integrality of $n^2_m$ is assured by the integrality of $n^0_m$.

  \vspace{.5cm}
  \noindent 
  $g=3$
  \vspace{.5cm}
  \\
  The relation between the topological amplitudes is given by
  \begin{equation}
  F_3^{\rm GV} = \frac{1}{ 32 \pi^4} \bar F_g^{\rm hol}.
  \end{equation}
  Going through  a similar analysis  for $g=0$
 it can be seen that the integrality of $n_m^3$ is ensured by 
 the integrality of $n^0_m$ and the  integrality of the 
 Fourier coefficient of the following  quasi-modular form
 \begin{eqnarray} \nonumber
 && \frac{1}{2}\left[\frac{1}{3024}-\frac{E_2E_4}{270\times 32}-\frac{E_6}{48\times 945}-\frac{E_2^3}{6\times 27\times 32}+\frac{1}{12\times 720}(6-5E_2^2-E_4)\right]  \\ 
 && \qquad \qquad  =4q^3+30q^2+120q^5+350q^6+840q^7+O(q^8).
 \end{eqnarray}
 In the second line we have performed the $q$-expansion and retained a few terms to 
 demonstrate that the Fourier coefficients are integers.

 \subsection{Conifold singularities}
 
 In the heterotic description, conifold singularities are points in moduli space
 where vectors becomes massless.  This will be reflected by singularities in the
 gravitational threshold. 
 Given the topological amplitude $\bar F_g^{\rm hol}$, we can determine 
 where the function becomes singular. 
 Note that the poly-logarithm function ${\rm Li}_a(z)$ is singular at $z=1$. 
The leading pole is  
for $a<0$
\be
{\rm Li}_a(z)=(z\frac{d}{dz})^{|a|}\left. \frac{1}{1-z}\right|_{z\rightarrow 1} \sim
 a!\frac{1}{(1-z)^{|a|+1}}.
\ee
Therefore the leading divergence for $g>1$ of $\bar F_g^{\rm hol}$,
is given by 
\be\label{conising}
\bar F_g^{\rm hol}|_{m\cdot y \rightarrow 0} \sim 
\frac{(-1)^{g-1}}{\pi^2}
\sum_{s=0}^{N-1}
e^{-2\pi i n_2s/N}c^{(r,s)}_{g-1}(m^2/2,0)(2g-3)!\frac{1}{\{1-e^{2\pi i m\cdot y}\}^{2g-2}}.
\ee
Note that we have removed the sum over the lattice point, since we are looking 
at lattice points where $m\cdot y =0$. 
This requirement  results in the equations 
\be
n_1T_1+n_2U_1=0,\qquad n_1T_2+n_2U_2=0.
\ee
Since $T_2, U_2>0$ and $m = ( n_1, n_2) >0$   we must have $n_2<0$. 
Therefore in the untwisted sector a possible  singularity  lies at
\begin{equation}
m  = ( 1, -1) , \qquad m ^2/2 = -1
\end{equation}
and in the twisted sector they lie at 
\begin{equation}\label{ptsconi}
m = (  \frac{r}{N} , -n_2) , \qquad r, n_2 >0 , \; \; \frac{m^2}{2} = - \frac{rn_2}{N}  , \; rn_2 \leq N.
\end{equation}
However from (\ref{untwissi}) we see that the sum 
\begin{equation}
\sum_{s=0}^{N-1}
e^{2\pi i s/N}c^{(0,s)}_{g-1}(-1,0) =0.
\end{equation}
Therefore there is no conifold singularity from the untwisted sector. 
In the twisted sector  using (\ref{coni1}) and (\ref{g0coni}) we see that the sum 
\begin{equation}\label{stconi}
\sum_{s=0}^{N-1}
e^{2\pi i n_2s/N}c^{(r,s)}_{g-1}(-\frac{rn_2}{N},0) =  n^0_{(\frac{r}{N}, n_2)} \times
\frac{2g-1}{2 ( 2g) !} (-1)^g (2\pi)^g B_{2g}.
\end{equation}
Now substituting  this result into (\ref{conising})  we obtain 
\begin{eqnarray}
\bar F_g^{\rm hol}|_{m\cdot y \rightarrow 0} 
 &\sim& -\frac{1}{2\pi^2} ( 2\pi)^g \frac{B_{2g}}{ 2g(2g-2)} \times
n^0_{(\frac{r}{N}, - n_2)},  \\ \nonumber
&=& \frac{1}{2\pi^2} ( 2\pi)^g \chi({\cal M}_g) \times
n^0_{(\frac{r}{N}, - n_2)}.
\end{eqnarray}
In the second line we have used the definition of the Euler characteristic of the 
moduli space of  genus $g$ Riemann  surfaces. 

Thus these models illustrate that  the strength of the conifold singularity is  also 
determined by the genus zero Gopakumar-Vafa invariant at that lattice point. 
For the un-orbifolded $K3\times T^2$ compactification studied in \cite{Marino:1998pg}, 
this is not transparent. 
This is because,  for this case the conifold singularity lies at the lattice point
$m = (1, -1)$ and $n^0_{(1, -1)} = -2$. 

There is one more observation of interest. 
From the relation  between the Gopakumar-Vafa form of the topological amplitude
and that obtained from the threshold integral in (\ref{relfgfhol}) together with (\ref{stconi})
 We  can show  that at the lattice points  corresponding to the conifold singularity 
 given in (\ref{ptsconi}) all the 
 higher genus topological  Gopakumar-Vafa invariants vanish, that is
 \begin{equation}
 n^g_{ (r/N, -n_2) } = 0 , \qquad {\hbox{for}}, \; r, n_2, g>0\;\;  rn_2 \leq N.
 \end{equation}

\section{Conclusions}\label{conclusion}

The evaluation of  $F_g$ for the compactification of the $E_8\times E_8$ 
heterotic string on  $\mathbb{Z}_N$  orbifolds  of $K3\times T^2$ by $g'$ generalizes 
the computation first done in \cite{Marino:1998pg} for the unorbifolded model. 
The results of this paper  show  that it is  the twisted elliptic genus of 
$K3$ by $g'$ which forms the crucial input which determines 
$F_g$ in these models.  It is in this sense    $M_{24}$  symmetry 
plays a role in these compactifications. 
Our analysis  was restricted to 
the standard embedding. It will be interesting to generalize  these results
to nonstandard embeddings.  
Certain non-standard embeddings that 
were studied in \cite{Chattopadhyaya:2017ews}  can form
a starting point. 
Recently gauge threshold corrections and the $F_1$ gravitational threshold  for $K3\times T^2$ model with non-zero 3-form flux were evaluated \cite{Angelantonj:2016gkz}. 
It will be also  interesting to generalize this to 
the orbifolds considered here.

Another interesting detail worth studying is the property of the gravitational threshold $F_g$ 
given as the sum of (\ref{fresultfg}) and (\ref{fresultdeg})  across walls of marginal stability. 
In \cite{Marino:1998pg}, it was shown that upto genus five, the corresponding gravitational threshold
for the $K3\times T^2$ model 
was a differentiable function across these walls. This involved details of the 
expansion coefficients
$c_{g-1}(l, t)$. It will be interesting to repeat this analysis for the models considered here. 

What is perhaps more interesting regarding the oribfolds  of 
$K3\times T^2$ inspired by $M_{24}$ symmetry is 
the prediction of  dual Calabi-Yau geometries on the type II A side. 
The analysis in this paper predicts the Euler numbers for each of these 
$K3$ fibered Calabi-Yau geometries.
It  also predicts the Gopakumar-Vafa invariants for the limit when the base $T^2$ of these
geometries is large. 
It will be interesting to explicitly construct these Calabi-Yau. 
In this context it will be interesting to look at freely acting 
quotients of Calabi-Yau  geometries such as that considered in  \cite{Braun:2010vc}
as well as Calabi-Yau  obtained by  non-geometric $K3$ fibrations in \cite{Hull:2017llx}. 

\acknowledgments

We thank Abhishek Chowdhury, Rajesh Gopakumar and Abhiram Kidambi  for discussions. 
A.C. thanks the Council of Scientific and Industrial Research (CSIR) for funding this project.

\appendix
\section{Modular transformations} \label{modt}

Modular properties result from the 
the basic formula for Poisson resummation  which is given by
\be
\sum_{n=-\infty}^{\infty}f(n)=\sum_{m =-\infty}^{\infty}\int_{\infty}^{\infty}dy \,f(y)e^{2\pi i m y}.
\ee

Using the above formulae on $m_1,m_2$ we arrive at 
\be \label{prft}
\sum_{m_1, m_2, n_1, n_2}p_R^{2k} q^{\frac{|p_L|^2}{2}}\bar{q}^{\frac{|p_R|^2}{2}}=\sum_{n_1,k_1,n_2,k_2}(\frac{-2\pi T_2}{\tau_2 \sqrt{2T_2U_2}})^{2k} (n_1\tau+k_1+U(n_2\tau+k_2))^{2k}e^{{\cal G}},
\ee
where ${\cal{G}}$ is given by 
\be
{\cal{G}}=-\frac{\pi T_2}{U_2\tau_2}|{\cal A}|^2-2\pi i T \det(A),
\ee
where 
\begin{eqnarray}
A=\left(\begin{matrix}n_1 & k_1 \\ n_2 & k_2  \end{matrix}\right)\\
{\cal A}=\left(\begin{matrix}1 & U  \end{matrix}\right)A\left(\begin{matrix}\tau\\1  \end{matrix}\right).
\end{eqnarray}

Using the Fourier transformation formulae one can now do the Poisson re-summation on all the four variables $m_1,m_2,n_1,n_2$ in each of the $(r,s)$ sectors and arrive at the following results:
\begin{eqnarray} \label{latticemod}
\Gamma_{2,2}^{(0,0)}(-\frac{1}{\tau},-\frac{1}{\bar{\tau}})&=&{\tau \bar{\tau}}\Gamma_{2,2}^{(0,0)}(\tau,\bar{\tau}),\\ \nonumber
\Gamma_{2,2}^{(0,s)}(-\frac{1}{\tau},-\frac{1}{\bar{\tau}})&=&{\tau \bar{\tau}}\Gamma_{2,2}^{(s,0)}(\tau,\bar{\tau}),\\ \nonumber
\Gamma_{2,2}^{(r,s)}(-\frac{1}{\tau},-\frac{1}{\bar{\tau}})&=&{\tau \bar{\tau}}\Gamma_{2,2}^{(s,N-r)}(\tau,\bar{\tau}).
\end{eqnarray}
Now changing the variable $\tau\rightarrow -\frac{1}{\tau}$ and $\bar{\tau}\rightarrow -\frac{1}{\bar{\tau}}$ in \ref{prft} and performing a Poisson resummation  where we take all the possible values of $n_1,n_2,k_1,k_2$ from $-\infty$ to $+\infty$) one can write the expression as
\begin{eqnarray}
&&\sum_{n_1,k_1,n_2,k_2 \in \mathbb{Z}}e^{{\cal G}(\frac{-1}{\tau},\frac{-1}{\bar{\tau}})}(\frac{-2\pi T_2 \tau\bar{\tau}}{\tau_2 \sqrt{2T_2U_2}})^{2k} (\frac{-n_1}{\tau}+k_1+U(-\frac{n_2}{\tau}+k_2)),\\ \nn
&&=\sum_{n_1,k_1,n_2,k_2 \in \mathbb{Z}}e^{{\cal G}(\tau,\bar{\tau})}(\frac{-2\pi T_2}{\tau_2 \sqrt{2T_2U_2}})^{2k} (-n_1\tau+k_1+U(-n_2\tau+k_2))\frac{(\tau\bar{\tau})^{2k+1}}{\tau^{2k}},\\ \nn
&&=\sum_{n_1,k_1,n_2,k_2 \in \mathbb{Z}}p_R^{2k}e^{{\cal G}}\frac{(\tau\bar{\tau})^{2k+1}}{\tau^{2k}}
\end{eqnarray}

Now assuming that $n_1,k_1$ are instead shifted from integers by $r/N$ and $ s/N$ the roles of $n_1, k_1$ would be reversed in the re-summed result, thus the shifts would become $s/N$ for $n_1$ and $(N-r)/N$ for $k_1$ as expected. Thus we have,
\begin{eqnarray}\nn
&& \sum_{\begin{smallmatrix}
	n_1 \in \mathbb{Z}+r/N,\\
	k_1 \in \mathbb{Z}+s/N,\\
	n_2,k_2 \in \mathbb{Z}\end{smallmatrix}}e^{{\cal G}(\frac{-1}{\tau},\frac{-1}{\bar{\tau}})}(\frac{-2\pi T_2 \tau\bar{\tau}}{\tau_2 \sqrt{2T_2U_2}})^{2k} (\frac{-n_1}{\tau}+k_1+U(-\frac{n_2}{\tau}+k_2))\\ \nn
&&=\sum_{\begin{smallmatrix}
n_1 \in \mathbb{Z}+s/N,\\
k_1 \in \mathbb{Z}+(N-r)/N,\\
n_2,k_2 \in \mathbb{Z}
	\end{smallmatrix}}e^{{\cal G}(\tau,\bar{\tau})}(\frac{-2\pi T_2}{\tau_2 \sqrt{2T_2U_2}})^{2k} (-n_1\tau+k_1+U(-n_2\tau+k_2))\frac{(\tau\bar{\tau})^{2k+1}}{\tau^{2k}}.\\
\end{eqnarray}
The above equations lead to,
\begin{eqnarray} \label{prmodtrans}
&&\sum_{m_1,m_2,n_1,n_2\in \mathbb{Z}}(p_R^{(0,0)})^{2k}e^{-\frac{2\pi i}{\tau}\frac{|p_L|^2}{2}}e^{\frac{2\pi i}{ \bar{\tau}}\frac{|p_R|^2}{2}}={\tau} \bar{\tau}^{2k+1}\sum_{m_1,m_2,n_1,n_2 \in \mathbb{Z}}(p_R^{(0,0)})^{2k}q^{\frac{|p_L|^2}{2}}\bar{q}^{\frac{|p_R|^2}{2}},\\
&& \sum_{m_1,m_2,n_1,n_2\in \mathbb{Z}}(p_R^{(0,s)})^{2k}e^{2\pi i m_1s/N}e^{-\frac{2\pi i}{\tau}\frac{|p_L|^2}{2}}e^{\frac{2\pi i}{ \bar{\tau}}\frac{|p_R|^2}{2}}={\tau} \bar{\tau}^{2k+1}\sum_{\begin{smallmatrix}
m_1,m_2,n_2 \in \mathbb{Z},\\ \nn
n_1\in \mathbb{Z}+r/N
\end{smallmatrix}}(p_R^{(s,0)})^{2k}q^{\frac{|p_L|^2}{2}}\bar{q}^{\frac{|p_R|^2}{2}},\\ \nn
&& \sum_{\begin{smallmatrix}m_2,n_2\in \mathbb{Z}\\
m_1\in \mathbb{Z}+s/N\\
n_1\in \mathbb{Z}+r/N\end{smallmatrix}}(p_R^{(r,s)})^{2k}e^{2\pi i m_1s/N}e^{-\frac{2\pi i}{\tau}\frac{|p_L|^2}{2}}e^{\frac{2\pi i}{ \bar{\tau}}\frac{|p_R|^2}{2}}={\tau} \bar{\tau}^{2k+1}\sum_{\begin{smallmatrix}m_1,m_2,n_2\in \mathbb{Z}\\
n_1\in \mathbb{Z}+s/N\end{smallmatrix}}(p_R^{(s,N-r)})^{2k}e^{2\pi i m_1(N-r)/N}q^{\frac{|p_L|^2}{2}}\bar{q}^{\frac{|p_R|^2}{2}}.
\end{eqnarray}
Here $p_R^{(r,s)}$ refers to 
\be
p_R^{(r,s)}=(\frac{-2\pi T_2}{\tau_2 \sqrt{2T_2U_2}}) (n_1\tau+k_1+U(n_2\tau+k_2)),
\ee
where $n_1\in \mathbb{Z}+r/N,\; k_1\in \mathbb{Z}+s/N,\; n_2,\; m_2\in \mathbb{Z} $.

Note that these are the  modular properties of lattice sectors that  are required 
to ensure that the integrand involved in evaluating the gravitational threshold is 
an invariant function under $SL(2, \mathbb{Z})$ transformations.

\section{Details for  $g=1$ threshold integral} \label{detailg1}

As discussed in  section \ref{genusone}, the one loop integral can be reduced to the 
integral given in (\ref{genusoneint}). 
This then receives contribution from  three inequivalent orbits which are given by 
\begin{enumerate}\label{appengenusone}
	\item  The zero orbit, ie, $A=0$,
	\item The non-degenerate orbit, ie, $A=\left(\begin{matrix}n_1 &j\\ 0 & p \end{matrix}\right), 
	\;\; p\neq0, \in \mathbb{Z}, \;\; n_1, j\in \frac{\mathbb{Z}}{N}, n_1>j $. 
	\item  The degenerate orbit, $A=\left(\begin{matrix}0 & j \\ 0 & p  \end{matrix}\right), 
	j \in \frac{\mathbb{Z}}{N}, p \in \mathbb{Z}$. 
\end{enumerate}
We write the contributions as 
\begin{eqnarray}
{\cal I } &=& \sum_A^{\prime}  \int_{{\cal F}_A} 
\frac{d^2 \tau}{\tau_2^2} T_2 
\exp\left( -\frac{\pi T_2}{U_2 \tau_2} |{\cal A}|^2 - 2\pi i T{\rm det} A\right) 
f^{(r, s) }( \tau) \hat E_2 ( \tau) , \\ \nonumber
&=& {\cal I}^{\rm zero} + {\cal I}^{\rm nondegen} + {\cal I}^{\rm degen}.
\end{eqnarray}

We now discuss in detail the contribution from each of the orbits.

\noindent
{\bf Zero Orbit}:  For $A=0$ we have $(r, s)= (0,0)$. 
Now note that $f^{(0,0)} \times \hat E_2$ is an almost modular form of weight zero. 
The integral is over the fundamental domain. Then the only contribution to the 
integral comes from 

 infinity \cite{Lerche:1988np}.
The result for integrals of this from  is given by the formula
\be
\int_{{\cal F}}\frac{d^2\tau}{\tau_2^2}F(q)\hat{G}_2^k(q)=
\left. \frac{1}{\pi (k+1)}G_2^{k+1}(q)F(q)\right|_{q^0},
\ee
where 
$\hat{G}_2(q)=2\zeta(2)(E_2-\frac{3}{\pi\tau_2})$.
This gives the first term of \ref{result} which is $\left. \frac{\pi}{6}T_2 E_2^2 f^{(0,0)}\right|_{q^0}$. 
Using the explicit expressions for $f^{(0, 0}$ for each of the orbifolds considered 
in this paper we obtain
\be
 \left. \frac{\pi}{6}T_2 E_2^2 f^{(0,0)} (q)\right|_{q^0}  = -48\pi \frac{T_2}{N}.
 \ee
Note that the analysis is done based on $T_2>U_2$ limit.
Therefore we have 
\begin{equation}
{\cal I}^{\rm zero} = \left. \frac{\pi}{6}T_2 E_2^2 f^{(0,0)} (q)\right|_{q^0}  = -48\pi \frac{T_2}{N}.
\end{equation}
\\

\noindent
{\bf Non-degenerate Orbit}: 
For the non-degenerate orbit we apply the unfolding technique, choose a representative of the matrix $A$ and integrate over two copies of the upper half plane. In this case $A$ becomes
\be
A=\left(\begin{matrix}n_1 & j \\ 0 & p  \end{matrix}\right), \qquad 
p >0, \in \mathbb{Z}, \;\; n_1, j\in \frac{\mathbb{Z}}{N}, n_1>j \geq 0 . 
\ee
The restriction to positive values in $A$ comes due to the fact that the 
negative values contribute equally and this is taken care of by the two copies of 
the upper half plane. 
We have  $\det (A)=n_1p$ and ${\cal A}=n_1\tau+j+pU$, where $p$ is non-zero.
The integration domain ${\cal F}_A$ is given by 
\be
-\infty< \tau_1<\infty, \qquad 0\le \tau_2<\infty.
\ee

In the (0,0) sector one has
and $j<n_1,\; j=0,1,...n_1-1$ and $p \in \mathbb{Z}$.
So we can proceed for the $\tau_1$ integral as done in \cite{Harvey:1995fq} 
for the un-orbifolded $K3$.  We write $\tau' _1=j+pU_1+n_1\tau_1$, then the relevant part for the $\tau_1$ integration in (\ref{appengenusone}) becomes
\begin{equation}\label{inttau1}
{\hat I(\tau_1)}=T_2\int \frac{d^2\tau}{\tau_2^{2}} 
{\rm exp}\left\{2\pi i n \tau_1-\frac{\pi T_2}{U_2\tau_2}\tau_1'^2\right\}\left(
\tilde{c}^{(0,0)}(n)-\frac{3}{\pi\tau_2}c^{(0,0)}(n)\right).
\end{equation}
Here we have focused on the $\tau_1$ dependent terms, and kept a generic term using the 
Fourier expansion of $f^{(0, 0} (q) \hat E_2(\tau) $. 

Replacing $\tau_1$ by $\tau_1'$
we see that the only $j$ dependence comes from 
\be \label{sumform}
e^{2\pi i n \frac{1}{n_1}(\tau_1'-j-pU_1)}.
\ee 
Now  performing the sum over $j$ forces $n_1 n_2=n$ where $n_2$ is an integer. 
Rest of the $\tau_1$ integration in the untwisted sector is Gaussian and the result is given by,
\begin{eqnarray} \nn
T_2\int_{-\infty}^{\infty} \frac{d\tau_1'}{\tau_2^{2}} 
{\rm exp}\left\{2\pi i n_2 \tau_1'-\frac{\pi T_2}{U_2\tau_2}\tau_1'^2\right\}\left(
\tilde{c}^{(0,0)}(n_1n_2)-\frac{3}{\pi\tau_2}c^{(0,0)}(n_1n_2)\right), \\ 
=\frac{\sqrt{T_2 U_2 \tau_2}}{\tau_2^2}(\tilde{c}^{(0,0)}(n_1n_2)-\frac{3}{\pi\tau_2}c^{(0,0)}(n_1n_2)) e^{-\pi n_2^2\tau_2 U_2/T_2} ,\\ \nn
=\frac{\sqrt{T_2 U_2 \tau_2}}{\tau_2^2}(\tilde{c}^{(0,0)}(m^2/2)-\frac{3}{\pi\tau_2}c^{(0,0)}(m^2/2)) e^{-\pi n_2^2\tau_2 U_2/T_2}.
\end{eqnarray}
If we are in  the untwisted sectors $(0,s)$,  we have $j=j'+s/N$ with $j'$ being an integer the above result holds but with an additional factor of $e^{-2\pi i n_2 s/N}$ and replacing $c^{(0,0)}(n_1n_2)$ and $\tilde{c}^{(0,0)}(n_1n_2)$ by $c^{(0,s)}(n_1n_2)$ and $\tilde{c}^{(0,s)}(n_1n_2)$.

Let us now look at the $\tau_2$ integration. The $\tau_2$ integrand is given by,
\be
{\cal I}_2=\sqrt{T_2 U_2}e^{-2\pi i n_2 s/N}\int_{0}^{\infty}\frac{d\tau_2}{\tau_2^{3/2}}\left(
\tilde{c}^{(0,s)}(m^2/2)-\frac{3}{\pi\tau_2}c^{(0,s)}(m^2/2)\right)\exp\{{\cal F}\},
\ee
where ${\cal F}$ is given by,
\be
{\cal F}=-2\pi \tau_2 n_2 n_1-\frac{\pi T_2}{U_2\tau_2}(n_1\tau_2+pU_2)^2-2\pi i T n_1 p-2\pi p n_2 U_1-\frac{\pi n_2^2 U_2\tau_2}{T_2}.
\ee
The $\tau_2$ integration is of the Bessel form and we use the following formulae to evaluate it,
\begin{eqnarray}
\int \frac{dx}{x^{3/2}}e^{-ax+b/x}&=&\sqrt{\pi/b}e^{-2\sqrt{ab}},\\ \nn
\int \frac{dx}{x^{5/2}}e^{-ax+b/x}&=&\sqrt{\pi}e^{-2\sqrt{ab}}\left(\frac{1+2\sqrt{ab}}{2b^{3/2}}\right).
\end{eqnarray}
The result of this exercise  yields for the untwisted sectors 
\begin{eqnarray} \label{appendtwist}\nn 
& & {\cal I}^{\rm nondegen}|_{\rm untwisted} =  
  4{\rm Re}  \sum_{\begin{smallmatrix}
		n_1> 0,n_2\geq-1\\n_1n_2\ne 0\\
		n_1\in \mathbb{Z}
\end{smallmatrix}} \left[
\tilde{c}^{(0,s)}(m^2/2)e^{-2\pi i sn_2/N}{\rm Li}_1(e^{2\pi i (n_1T+n_2U)}) \right. 
\\ \nonumber 
& & \qquad\qquad\qquad\qquad \left. -\frac{3}{\pi T_2U_2}c^{(0,s)}(m^2/2)e^{-2\pi i sn_2/N}{\tilde{\cal{P}}}(n_1T+n_2U) \right],
\end{eqnarray}
where $\tilde{\cal P}$ is given by
\be
\tilde{\cal P}(x)={\rm Im}(x){\rm Li}_2(e^{2\pi i x})+\frac{1}{2\pi}{\rm Li}_3(e^{2\pi i x}).
\ee

Let us go over to the twisted sectors. 
First  consider an order $N$ orbifold when $N$ is prime. 
 In the twisted sectors,  the $n$ in the  Fourier coefficients $c^{(r, s)}(n)$ and 
 $\tilde c^{(r, s) } (n)$  is such that $n \in \mathbb{Z}/N$. 
 While
 $j=j'+s/N,\; j'\in \mathbb{Z}$ and  $n_1 \in r/N+\mathbb{Z}$.
The modular forms $f^{(r,s)}$ has the property that  the transformations of  $\tau\rightarrow \tau+1$ 
can be used to relate it to 
 $f^{(r,0)}$.  
  
 After doing this, one effectively works in the $(r, 0)$ twisted sector but now $j$ runs from
  $0$ to $n_1-1/N$ in steps of $1/N$ \footnote{See \cite{David:2006ji} for a discussion
  of this step for integrands which involve a $\Gamma_0(2)$ from.}. 
  Thus from (\ref{sumform}) we see that 
   the sum over all the values of $j$  for a prime $N$ would be $Nn_1$ iff $n/(n_1N) =n_2$ is an integer, else it is zero.   Therefore $n = n_1 n_2 N$,  only  integral values of $n$ are picked
   up. 
However one observes that the coefficients of $q^{n}$ in $f^{(r,s)}$ are related to the coefficients of $q^{n}$ in $f^{(r,0)}$ 
  under  transformation $\tau \rightarrow \tau+1$. If  $n$ is fractional the relation is 
  through a multiplicative  phase.
Using this property one can write the result in terms of the Fourier coefficients in the  $f^{(r,s)}$
sector.

The final result for a prime $N$ for the non-degenerate orbit can be given by
\begin{eqnarray}\label{appendtwist1} \nn
& & {\cal I}^{\rm nondegen}|_{\rm twisted} = 
4{\rm Re} \sum_{\begin{smallmatrix}
	n_1\> 0,n_2\geq-1\\ n_1n_2\ne 0\\
	n_1\in \mathbb{Z}+r/N
	\end{smallmatrix}} \left[
	\tilde{c}^{(r,s)}(m^2/2)e^{-2\pi i sn_2/N}{\rm Li}_1(e^{2\pi i (n_1T+n_2U)}) 
	\right. 
	\\ 
& &\qquad\qquad\qquad\qquad
 \left. 	-\frac{3}{\pi T_2U_2}c^{(r,s)}(m^2/2)e^{-2\pi i sn_2/N}{\tilde{\cal{P}}}(n_1T+n_2U) \right].
\end{eqnarray}

The above argument also holds when 
when $N$ is composite  and the twisted sector  $r$ is such that $r,N$ are co-prime.
However if $r$ divides $N$, then there are various sub-orbits for the sectors $(r, s)$
under transformations $\tau \rightarrow \tau +1$. 
For example if $N=4$, then      the sectors $(2, 0), (2, 2) $ and the sectors $(2, 1), (2, 3)$ 
form distinct sub-sectors not related by transformations of the kind $\tau \rightarrow \tau+1$. 
One then uses the argument developed for $N$ prime and carries it out for each 
of the sub-orbits.  The end result  works out to be  the same as that given in 
(\ref{appendtwist1}). 

\noindent
{\bf Degenerate Orbit}: 
Here the determinant of $A$ is 0. So we choose the
matrix $A$ as
\be \nn
A=\left(\begin{matrix}0 & j \\ 0 & p  \end{matrix}\right),
\ee
where $j \in \mathbb{Z}/N, p \in \mathbb{Z}$ and $(j,p)\neq (0,0)$. Since $n_1=0$ twisted sectors don't contribute to the degenerate orbit. There are logarithmic divergences due to the constant
term in the expansion of $f^{(0,s)}E_2(\tau)$ which needs to be renormalized. We have
\begin{eqnarray}\nn
\hat {\cal I }=\sum_{A{^{\prime}},n, s} \int_{{\cal F}_A} 
\frac{d^2 \tau}{\tau_2^2} T_2 
\exp\left( -\frac{\pi T_2}{U_2 \tau_2} |{\cal A}|^2 \right) \left(
\tilde{c}^{(0,s)}(n)-\frac{3}{\pi\tau_2}c^{(0,s)}(n)\right)q^n ,
\end{eqnarray}
where $ |{\cal A}|^2=|j+pU|^2$, $p\in \mathbb{Z},\; j\in \mathbb{Z}+s/N$ in $(0,s)$ sector.
To  regularize the result 
 we multiply the integrand for the $\tilde{c}^{(0,s)}$ with a 
 factor of $(1-e^{-\Lambda/\tau_2})$.

The integration domain ${\cal F}_A$  in this orbit is given by 
\be
-1/2\le \tau_1<1/2, \qquad 0\le \tau_2<\infty.
\ee
The only dependence of $\tau_1$ comes from the $q^n$ and since it is integrated from $-1/2\le \tau_1<1/2$ it picks up only the $q^0$ term from $f^{(0,s)}\hat{E}_2$.
Performing the  $\tau_2$ integration for the  term containing the coefficient 
$\tilde{c}^{(0,s)}$ we obtain
\be \label{sum1}
\tilde{c}^{(0,s)}(0)\left[\frac{U_2}{\pi} \left( 
 \sum_{j,p}\frac{1}{|j+pU|^2}-\frac{1}{|j+pU|^2+\Lambda U_2/(\pi T_2)} \right) 
 -\int_{{\cal F}_A}\frac{d\tau_2}{\tau_2}(1-e^{-\Lambda/\tau_2}) \right].
\ee
The result of the integral from the  second term is given by 
\be\label{sum3}
-\tilde{c}^{(0,s)}(0)\left(\log(\Lambda)+\gamma_E+1+\log (2/3\sqrt{3})\right).
\ee
Now for  coefficient of ${c}^{(0,s)}$ one gets  after integrating $\tau_2$ the result
\be\label{sum2}
-\sum_{j,p}\frac{6}{\pi^3}\frac{U_2^2}{T_2}{c}^{(0,s)} \frac{1}{|j+pU|^4}.
\ee
Note that here $j = j' + s/N$ where $j'\in \mathbb{Z}$. This is the difference as compared
to the calculations of \cite{Harvey:1995fq}. 
 To evaluate these sums on $j,p$ we need to divide the sum in $j'\ne 0,\; p=0$ and $p>0,\; j'\in \mathbb{Z}$ and use the following results \cite{Harvey:1995fq}
\begin{eqnarray} \label{sum}
\sum_{j'\in \mathbb{Z}}\frac{1}{(j'+B)^2+C^2}&=& \frac{\pi}{C}\left[1+\frac{e^{2\pi i (B+iC)}}{1-e^{2\pi i (B+iC)}}+\frac{e^{-2\pi i (B-iC)}}{1-e^{-2\pi i (B-iC)}}\right]\\  \label{sumder}
-\frac{1}{2C}\partial_C\sum_{j'\in \mathbb{Z}}\frac{1}{(j'+B)^2+C^2}&=&\sum_{j\in \mathbb{Z}}\frac{1}{((j'+B)^2+C^2)^2}.
\end{eqnarray}
Note that in applying these formula, the information of $s$ is present in $B$ as now 
$B = U_1p + s/N$. 
When $p=0$, the terms in the sum  that  contribute in the $\Lambda\rightarrow \infty $ limit 
from (\ref{sum1}) and (\ref{sum2}) 
result in 
\be\label{a11}
\sum_{s=0}^{N-1} \left( 
\tilde{c}^{(0,s)}(0)\frac{2\zeta(2,s/N)}{\pi}U_2-{c}^{(0,s)}(0)\frac{6\zeta(4,s/N)}{\pi^3} \frac{U_2^2}{T_2} \right) .
\ee
Now we carry out the sum over $j'$ for $p\neq 0$  in (\ref{sum2}) using (\ref{sumder}). 
When one does this, there is a term that results from the 
action of the derivative on the first term of (\ref{sum}), that is the term independent of $B$.
This results in 
\be \label{a12}
\sum_{s=0}^{N-1}\frac{6}{\pi^2}c^{(0,s)}(0)\frac{\zeta(3)}{T_2U_2}.
\ee
Now, lets carry out the sum over $j'$  for $p\neq 0$ in the first two terms of  (\ref{sum1}). 
Applying (\ref{sum}) from   the $B$ independent part we get 
\be
\sum_{p=1}^\infty
\left( \frac{2}{p}-\frac{2}{\sqrt{p^2+\Lambda/\pi T_2U_2}}=-\ln \frac{\pi T_2U_2}{\Lambda}-\ln 4+2\gamma_E \right) .
\ee
This again comes in all the sectors of $(0,s)$ and taking care of the moduli dependence 
and combining (\ref{sum3}) we tobain
\be \label{a13}
\sum_{s=1}^{N-1} \tilde{c}^{(0,s)}(-\log T_2U_2-\kappa),
\ee
where $\kappa=1-\gamma_E+\log(\frac{8\pi}{3\sqrt{3}})$.

Finally keeping track of all 
 the second and third terms in the  RHS of \ref{sum} and its derivative \ref{sumder}, that
 is the $B$ dependent part both in the sums (\ref{sum1}) and (\ref{sum2}) 
we obtain
\be\label{a14}
4{\rm Re}\sum_{\begin{smallmatrix}
		l\ne 0,l\geq-1\\
		l\in \mathbb{Z}
\end{smallmatrix}} \left[
\tilde{c}^{(0,s)}(0)e^{2\pi i sl/N}{\rm Li}_1(e^{2\pi i (lU)})-\frac{3}{\pi T_2U_2}c^{(0,s)}(0)e^{2\pi i sl/N}{\tilde{\cal{P}}}(lU) \right].
\ee

Combining all these results 
\begin{equation}
{\cal I}^{degen} =  (\ref{a11}) + ( \ref{a12}) + ( \ref{a13})+  ( \ref{a14}) .
\end{equation}

\section{Details for the $g>1$, threshold integral} \label{detailg2}

The integral we have to do is the one in (\ref{fgexp2}). 
As discussed  there is no zero orbit.  We have to perform the integral 
over the non-degenerate orbit   and the 
degenerate orbit.

\noindent
{\bf Non-degenerate orbit:}
This orbit are characterized by matrices of the form 
For the non-degenerate orbit we apply the unfolding technique and choose a representative of the matrix $A$ and integrate over two copies of the upper half plane. In this case $A$ becomes
\be
A=\left(\begin{matrix}n_1 & j \\ 0 & p  \end{matrix}\right), \qquad
p\in \mathbb{Z} , n_1, j \in \frac{\mathbb{Z}}{N}, n_1\geq j \geq 0.
\ee
This gives $\det A=n_1p$ and ${\cal A}=n_1\tau+j+pU$.
The domain for the integration now is two copies of the upper half plane.
In the $(0,s)$ sectors and $(r,s)$ sectors one can verify that the limits on $j,n_1$ are as in the last section for $g=1$. 
The $j$ dependence is in the phase given in (\ref{sumform}). 
 Again after shifting $\tau_1$ to $\tau' _1=j+pU_1+n_1\tau_1$ which forces  the 
 $n$ of the Fourier coefficient to be integer as before. 

The relevant part for the $\tau_1$ integration in (\ref{fgexp2}) becomes
\begin{eqnarray}\label{inttau1g}
{I(\tau_1)}&=&T_2\int \frac{d^2\tau}{\tau_2^{2+t}} \left(\frac{T_2}{2U_2}\right)^k(\tau'_1+i (pU_2+n_1\tau_2))^{2k} \\ \nn
&& {\rm exp}\left\{2\pi i n \tau_1-\frac{\pi T_2}{U_2\tau_2}\tau_1'^2\right\}.
\end{eqnarray}

Note that the coefficients here would be written as $c^{(r,s)}(m^2/2,t)$ in different sectors and a factor of $e^{-2\pi i s n_2/N}$ will come as an extra phase to the integral in the corresponding sectors along with $e^{-2\pi i T n_1 p}$. Also a factor of 2 would be present due to the two copies of the upper half plane. We shall not write these in the integral to avoid cumbersome notation and will only reinstate them back in the final result.
Let us denote the vector $\vec m$ with the components 
\be
\vec m=(n_1, n_2), \qquad m^2/2=n_1n_2,
\ee
 where $n_2 \in\mathbb{Z}$ and   $n_1\in \mathbb{Z}+r/N$.

This integral  in (\ref{inttau1g}) 
can be done  by using binomial expansion for $(\tau'_1+i (pU_2+n_1\tau_2))^{2k}$ and shifting the variable $\tau_1'$ to $\tilde{\tau}_1 = \tau'_1-i n_2 U_2 \tau_2/{T_2}$.  
Following through these replacements in \ref{inttau1g}, we  obtain
\begin{eqnarray}
{I(\tau_1)}&=&T_2\int \frac{d^2\tau}{\tau_2^{2+t}} \left(\frac{T_2}{2U_2}\right)^k(\tilde{\tau}_1+in_2 U_2 \tau_2/T_2+i (pU_2+n_1\tau_2))^{2k} \\ \nn
&& {\rm exp}\left\{-\frac{\pi T_2}{U_2\tau_2}\tilde{\tau}_1^2-\frac{\pi U_2\tau_2 n_2^2}{T_2}\right\}.
\end{eqnarray}
The $\tilde{\tau}_1$ can be integrated and we are left with
\bea
{I(\tau_1)}&=& T_2\frac{1}{\tau_2^{2+t}}
\left(\frac{T_2}{2U_2} \right)^k  \sum_{h=0}^{2k}\sum_{j=0}^{[h/2]} \left(\begin{matrix}
	h \\
	2j
\end{matrix}\right)\left(\begin{matrix}
2k \\
h
\end{matrix}\right) (i n_2 U_2 \tau_2/T_2)^{h-2j} (ipU_2+in_1\tau_2)^{2k-h} \nn \\
&&\Gamma(j+1/2)\frac{1}{(\pi T_2/U_2 \tau_2)^{j+1/2}} e^{-\pi U_2 \tau_2 n_2^2/T_2}.
\eea
Here we have written down only the terms that involve $\tau_1$ dependence. 
Now putting back all the 
terms which have $\tau_2$ dpendence we 
expand the complete integral as a polynomial in $\tau_2$ which gives
\begin{eqnarray} \nn
\tilde{I}&=&\sum_{h=0}^{2k}\sum_{j=0}^{[h/2]}\sum_{l=0}^{2k-h}\left(\begin{matrix}
	h \\
	2j
\end{matrix}\right)\left(\begin{matrix}
2k \\
h
\end{matrix}\right)\left(\begin{matrix}
2k-h \\
l
\end{matrix}\right) T_2(\frac{T_2}{2U_2})^k e^{-2\pi i T_1 n_1 p}\Gamma(j+1/2)(-1)^{k-j} \\ \nn
&&
\int_{0}^{\infty} d\tau_2 \, (\tau_2)^{l+h-j-3/2-t}\; {\rm exp}[-\pi T_2 U_2 p^2/\tau_2-\pi \tau_2/(T_2 U_2)(n_1T_2+n_2 U_2)^2]\\ 
&& n_1^l (pU_2)^{2k-h-l}(n_2 U_2/T_2)^{h-2j}(U_2/(\pi T_2))^{j+1/2} \label{tau1int}.
\end{eqnarray}
Integrating  term by term and replacing $h$ by $2k-h$, which still  keeps the limits as $0$ to $2k$,
we obtain
\begin{eqnarray} \nn
&&\tilde{I}=\sum_{h=0}^{2k}\sum_{l=0}^{2k-h}\sum_{j=0}^{[k-h/2]}\left(\begin{matrix}
	2k-h \\
	2j
\end{matrix}\right)\left(\begin{matrix}
	2k \\
	h
\end{matrix}\right)\left(\begin{matrix}
	h \\
	l
\end{matrix}\right) T_2(\frac{T_2}{2U_2})^k e^{-2\pi i T_1 n_1 p}\Gamma(j+1/2)(-1)^{k-j} \\ \nn
&& n_1^l (pU_2)^{h-l}(n_2 U_2/T_2)^{2k-h-2j}(U_2/(\pi T_2))^{j+1/2} \\ \nn
&& 2 \left( \frac{|p|T_2 U_2}{n_1 T_2+n_2 U_2} \right)^{l+2k-h-j-t-1/2} 2 c^{(r,s)}(r^2/2, t)e^{-2\pi i s n_2/N}K_{l+2k-h-j-t-1/2}(2\pi i |p|(n_1T_2+n_2 U_2)).\\
\end{eqnarray}

The result of this integral seems manifestly different from that 
obtained in \cite{Marino:1998pg} at this step. 
To bring into in the from we perform changes in the order of summations and limits. 
We outline these now. 
 First we change the order of summation on $l$ and $h$ which gives
\be
\sum_{h=0}^{2k}\sum_{l=0}^{2k-h}\sum_{j=0}^{[k-h/2]}  \rightarrow \sum_{l=0}^{2k}\sum_{h=0}^{2k-l}\sum_{j=0}^{[k-h/2]}.
\ee
Now we call $h-l $ as $h^+$ and  change the orders of the sums on $h^+$ and $l$.
\be
 \sum_{l=0}^{2k}\sum_{h=0}^{2k-l}\sum_{j=0}^{[k-h/2]} \rightarrow
 \sum_{h^+=0}^{2k}\sum_{l=0}^{2k-h^+}\sum_{j=0}^{[k-(h^++l)/2]}.
\ee
Now one needs to change the order of the sums $j$ and $l$ which can be done as
\be
\sum_{h^+=0}^{2k}\sum_{l=0}^{2k-h^+}\sum_{j=0}^{[k-(h^++l)/2]} \rightarrow
\sum_{h^+=0}^{2k}\sum_{j=0}^{[k-h^+/2]} \sum_{l=0}^{2k-2j-h^+}.
\ee
The final sum over $l$  is doable and this gives us
\begin{eqnarray} \nn
\tilde{I}&=& 2^{1-k}\sum_{t=0}^{k+1}\sum_{n_1n_2\ne 0,p\neq 0}e^{-2\pi i s n_2/N} c^{(r,s)}(m^2/2, t)\sum_{h^+=0}^{2k}\sum_{j=0}^{[k-h^+/2]}\frac{(2k)!}{j! h^+!(2k-h^+-2j)!( 4\pi) ^j}\\ \nn
&&p^{h^+} (T_2U_2 )^{j+h^++1/2-k} (n_2 U_2+n_1 T_2)^{2k-h^+-2j} (-1)^{k-j} \left( \frac{|p|T_2 U_2}{n_1 T_2+n_2 U_2} \right)^{2k-h^+-j-t-1/2}\\ \nn
&& K_{2k-h^+-j-t-1/2}(2\pi |p|(n_1T_2+n_2 U_2)) e^{-2\pi i p(n_1T_2+n_2 U_2)}.\\ \label{resultkk}
\end{eqnarray}

The result agrees with  equation (4.29) of 
\cite{Marino:1998pg} when $K^3\times T^2$ is not orbifolded.  In (\ref{resultkk}), the summation
over the orbifold sectors is implied. 

To rewrite the result in terms of poly-logarithms we use the relations
\begin{eqnarray} \label{polylog}
K_{s+1/2}(x)&=&\sqrt{\frac{\pi}{2x}}e^{-x}\sum_{k=0}^{s}\frac{(s+k)!}{k!(s-k)!}\frac{1}{(2x)^k},\\
{\rm Li}_m(x)&=&\sum_{l=1}^\infty \frac{x^l}{l^m}. 
\end{eqnarray}
This leads to 
\begin{eqnarray} \nn
\tilde{I}&=& 2^{1-k}\sum_{t=0}^{k+1}\sum_{ n_1, n_2\neq 0} \sum_{h^+=0}^{2k}\sum_{j=0}^{[k-h^+/2]}\sum_{a=0}^s\frac{(2k)!}{j! h^+!(2k-h^+-2j)!}\frac{(s+a)!}{a!(s-a)!}\frac{(-1)^{k-j+h^+}}{( 4\pi) ^{j+a}}\\ \nn
&& (T_2U_2 )^{k-t} {\rm(sgn\;Im}(m\cdot y ))^{h^+}
c^{(r,s)}(m^2/2, t)e^{-2\pi i s n_2/N} {\rm Im}(m\hat{\cdot}y)^{t-j-a}{\rm Li}_{1+a+j+t-2k}(e^{2\pi i m\hat{\cdot}y}).\\ \label{resultli}
\end{eqnarray}

The case $m=0$ needs to be treated separately. 
From (\ref{inttau1g}) and (\ref{tau1int}) we see that we can perform the integral 
also for the limit $n_1 = n_2=0$. 
 After the $\tau_1$ integration, for the $m=0$ limit we obtain
\begin{eqnarray} \nn
\tilde{I}|_{m=0} &=&\sum_{j=0}^{k}\left(\begin{matrix}
	2k \\
	2j
\end{matrix}\right)
 T_2(\frac{T_2}{2U_2})^k \Gamma(j+1/2)(-1)^{k-j} \\ \nn
&&
\int_{0}^{\infty} d\tau_2 \, (\tau_2)^{j-3/2-t}\; {\rm exp}[-\pi T_2 U_2 p^2/\tau_2]  (pU_2)^{2k-2j}(U_2/(\pi T_2))^{j+1/2}. \\ \label{tau1r0}
\end{eqnarray}
One can  shift to the variable $j'= k-j$ which also runs from 0 to $k$.
We will re-label this variable $j$. 
The resultant integral is a gamma function and since there is 
 a sum over $p$ it forms a zeta function. A factor of 2  arises  on  taking two copies of the upper half plane. However for values  $t=k$ 
 the zeta function blows up so one can use a regularization 
 by adding $-\epsilon$ to the power of $\tau_2$.
We have
\be
\int e^{-\pi T_2 U_2 p^2/\tau_2} \tau_2^{k-j-t+1/2-\epsilon}\;\frac{d\tau_2}{\tau_2^2}=
\left(\frac{1}{\pi T_2 U_2 p^2}\right)^{j-k+t+1/2+\epsilon}\;\Gamma(j-k+t+1/2+\epsilon).
\ee
So from \ref{tau1r0} the contribution from  $ m =0$ to the 
 non-degenerate part of the integral is given by,
\begin{eqnarray} \label{ndegr0}
{\cal I}^{\rm nondegen}|_{ m=0}&=& 2\sum_{j=0}^k \sum_{t=0}^{k+1}(-1)^j 2^{2j-4k+t} \frac{(2k)!}{(2j)!(k-j)!}e^{-2\pi i s n_2/N}\left(\frac{c^{(0,s)}(0,t)}{\pi^{t+1/2}}\right) \nn \\
&&(\pi T_2 U_2)^{-\epsilon}\Gamma(1/2+j+t-k+\epsilon)\zeta(1+2\epsilon+2t-2k).
\end{eqnarray}
In the equation \ref{ndegr0} we have used the fact 
\be
\Gamma(n+1/2)=\frac{(2n)!\sqrt{\pi}}{n!4^n}.
\ee
We now use 
the  equation \ref{ndegr0}  and consider $\epsilon\rightarrow 0$ limit  to extract out the 
finite contribution as follows,
\bea \nn
(\pi T_2 U_2)^{-\epsilon}&=&1-\epsilon \log\left({\pi}{T_2U_2}\right),\\ \nn
\Gamma(1/2+s+\epsilon)&=&\Gamma(1/2+s)(1+\epsilon\, \psi (1/2+s)), \\
\zeta(1+2\epsilon)&=&\frac{1}{2\epsilon}+\gamma_E,
\eea
where $\psi(x)$ is the logarithmic derivative of the gamma function. 
The resulting  sums in \ref{ndegr0} can be explicitly computed for some cases.
We use the results $\Gamma(1/2+n)=\pi^{1/2}2^{-n}(2n+1)!!$ and $\Gamma(1/2-n)=(-1)^n\pi^{1/2}2^{n}/(2n-1)!!$ with $n\ge 0$ to arrive at
\be
\sum_{j=0}^k (-1)^j 2^j \frac{(2k!)}{(2j)!(k-j)!}\Gamma(1/2+j)=\pi^{1/2}\delta_{k,0}.
\ee
Thus results for $k=0$ that is  genus  one
 and $k\ne 0$ are  different. For $k> 0$ we can write the full non-degenerate orbit result as,
\bea
{\cal I}^{\rm nondegen}_{k> 0, m=0} &=&2\sum_{s=0}^{N-1} \sum_{t=0}^{k-1} \frac{c^{(0,s)}(m^2/2,t)}{\pi^{t+1/2}} \zeta(1+2(t-k))(2T_2U_2)^{(k-t)} \\ \nn
&& \sum_{\tilde s=0}^k (-1)^{\tilde s} 2^{2(\tilde s-2k)+t} \frac{(2k)!}{(2\tilde s)!(k-\tilde s)!}\Gamma(1/2+\tilde s+t-k) \\ \nn
&& + \sum_{s=0}^{N-1} \frac{c^{(0,s)}}{2^{3k}\,\pi^{k}} \sum_{\tilde s=0}^k (-1)^s \frac{(2k)!}{(2\tilde s)!(k-\tilde s)!} \psi(1/2+ \tilde s).
\eea

Putting all the contributions together 
the contribution for  $F_{g>1}^{\rm nondegen}$  is given by 
\bea \label{resndeg}
F_{g>1}^{\rm nondegen} 
&=& \frac{(-1)^{g-1}}{2^{2(g-1)}\pi^2}\sum_{r,s}\sum_{ m\ne 0}\sum_{t=0}^g \sum_{h=0}^{2g-2}\sum_{j=0}^{g-1-h/2}\sum_{a=0}^{\hat s} e^{-2\pi i s n_2/N}c^{(r,s)}(m^2/2,t)\\ \nn
&& \frac{(2g-2)!}{j!h!(2g-h-2j-2)!} \frac{(-1)^{j+h}}{(4\pi)^{j+a}}\frac{(\hat s+a)!}{(\hat s-a)!a!}({\rm sgn}({{\rm Im}} (m\cdot y)))^h\\ \nn
&&\frac{ ({\rm Im} (m\hat \cdot y) )^{t-j-a} }
{(T_2U_2)^t} {\rm Li}_{3+a+j+t-2g}(e^{2\pi i m\hat \cdot y})\\ \nn
&+& \sum_{s=0}^{N-1} \frac{c^{(0,s)}(0,g-1)}{2^{4g-5}\, \pi^{g+1}}\frac{1}{(T_2U_2)^{g-1}} \sum_{\tilde s=0}^{g-1}(-1)^s \frac{(2g-2)!}{\tilde s!(g-1-\tilde s)!}\psi (1/2+\tilde s)\\ \nn
&+& \sum_{s=0}^{N-1}\sum_{t=0}^{g-2} \frac{c^{(0,s)}(0,t)}{\pi^{t+5/2}} \, \frac{\zeta(3+2(t-g))}{(T_2U_2)^t}\\ \nn
&& \sum_{\tilde s=0}^{g-1}(-1)^{\tilde s} 2^{2(\tilde s-2g+2)}\frac{(2g-2)!}{(2\tilde s)!(g-1-\tilde s)!}\Gamma(3/2+\tilde s+t-g).
\eea
Here $ m \neq 0$  is defined in (\ref{limits}) and $\hat s$ is defined in (\ref{hatsdef}).

\noindent
{\bf Degenerate Orbit: }
The degenerate orbit is characterised by matrices of the 
form 
\begin{equation}
A = \left( \begin{array}{cc}
0 & j \\ 0 & 0 
\end{array} \right),  \qquad j \in \frac{\mathbb{Z}}{N} , \; j\neq 0 .
\end{equation}
 The evaluation of the integral in the degenerate orbit proceeds 
 as follows. 
 Let us first assume $T_2>U_2$ for this part of the calculation we have
\be\label{deg}
I^{deg}=\frac{T_2^{k+1}}{(2U_2)^k}\int j^{2k}\tau_2^{-t-2}d\tau_2 e^{-\pi j^2 T_2/U_2\tau_2-2\pi n'\tau_2}\int_{-1/2}^{1/2} d\tau_1 e^{2\pi in'\tau_1}.
\ee
Note that since we are in the un-twisted sector for this orbit $n' \in \mathbb{Z}$. 
Furthermore to 
obtain a non-zero result due to the integration we require $n'=0$. Then the result for 
$I^{\rm degen}$ is  given by
\be\label{degr}
I^{deg}=\sum_{s=0}^{N-1}\sum_t\pi^{-t-1}U_2 (T_2/U_2)^{k-t}c^{(0,s)}(0,t)t! \zeta(2(t+1-k,s/N)).
\ee
If $U_2>T_2$, we need to in fact   we need start by performing at T-duality and then 
we obtain \cite{Borcherds:1996uda,Harvey:1995fq}  \footnote{The reason this has to be 
be done has to do with the convergence of the integral for Fourier coefficient when $n'=-1$. }
\be
I^{\rm degen}=\sum_{s=0}^{N-1}\sum_t\pi^{-t-1}T_2 (U_2/T_2)^{k-t}c^{(0,s)}(0,t)t! \zeta(2(t+1-k), s/N).
\ee
Now reinstating all factors we obtain 
\be \label{resdeg}
F_{g>1}^{\rm degen}=\frac{1}{T_2^{2g-3}}\sum_{s=0}^{N-1}\sum_{t=0}^g c^{(0,s)}(0,t)\frac{t!}{2^{2g-2}\pi ^{t+3}}\left(\frac{T_2}{U_2}\right)^t\zeta(2(2+t-g),s/N).
\ee

\section{ Genus zero GV invariants for CHL orbifolds} \label{GVZ}

In this appendix we list the genus zero Gopakumar-Vafa invariants 
$n^0_m$ corresponding to the CHL orbifolds. 
We observe that they are integers. 
For the remaining orbifolds in table \ref{tt}, we provide an ancillary Mathematica
code from which these invariants can be evaluated and seen to be integers. 
One point worth mentioning, is that all the remaining orbifolds 
especially the case of $2B$ and $3B$ do not have any geometric 
interpretation as actions on the $K3$ surface. Therefore 
the fact that the Gopakumar-Vafa invariants turn out to be integers is 
interesting.

\begin{table}[H]
	\renewcommand{\arraystretch}{0.5}
	\begin{center}
		\vspace{0.5cm}
		\setlength{\doublerulesep}{20\arrayrulewidth}
		{\scriptsize{
				\begin{tabular}{|c|c|c|c|c|c|}
					\hline
					& & & & & \\
					$(n_1,n_2)$ & $(1,-1)$ & (1,0) & (1,1) & (1,2) & (1,3)  \\
					\hline
					& & & & & \\
					$n^0_{(n_1,n_2)}$ & $-2$ & 480 & 282888 & 17058560 & 477516780 \\
					\hline
					\hline
					& & & & & \\
					$(n_1,n_2)$  & (1,4) & (1,5) & (1,6) & (1,7) & (1,8) \\
					\hline
					& & & & & \\
					$n^0_{(n_1,n_2)}$ & 8606976768 & 115311621680 & 1242058447872 & 11292809553810  & 89550084115200   \\
					\hline
				\end{tabular}
		}}
		
	\end{center}
	\vspace{-0.2cm}
	\caption{$N=1$, $K3$ itself.}
	\renewcommand{\arraystretch}{0.5}
\end{table}

\begin{table}[H]
	\renewcommand{\arraystretch}{0.5}
	\begin{center}
		\vspace{0.5cm}
		\setlength{\doublerulesep}{20\arrayrulewidth}
		{\scriptsize{
				\begin{tabular}{|c|c|c|c|c|c|}
					\hline
					& & & & & \\
					$(n_1,n_2)$ & $(0,n_2)$ & (1,1) & (1,2) & (1,3) & (1,4)  \\
					\hline
					& & & & & \\
					$n^0_{(n_1,n_2)}$ & $-32$ &151552 & 8387328 &47890048 & 4294949632 \\
					\hline
					\hline
					& & & & & \\
					$(n_1,n_2)$ & $(1/2,0)$ & (1/2,2) & (1/2,4) & (1/2,6) & (1/2,8) \\
					\hline
					& & & & & \\
					$n^0_{(n_1,n_2)}$ & 512 & 151552 & 8671232 &240009216 &4312027136  \\
					\hline
					\hline
					& & & & & \\
					$(n_1,n_2)$ & $(1/2,-1)$ & (1/2,1) & (3/2,1) & (5/2,1) & (7/2,1) \\
					\hline
					& & & & & \\
					$n^0_{(n_1,n_2)}$ & 16 & 8128 & 1212576 & 47890048 & 1055720304 \\
					\hline
				\end{tabular}
		}}

	\end{center}
	\vspace{-0.2cm}
	\caption{$N=2$, 2A orbifold.}
	\renewcommand{\arraystretch}{0.5}
\end{table}

\begin{table}[H]
	\renewcommand{\arraystretch}{0.5}
	\begin{center}
		\vspace{0.5cm}
		\setlength{\doublerulesep}{20\arrayrulewidth}
		{\scriptsize{
				\begin{tabular}{|c|c|c|c|c|c|}
					
					\hline
					& & & & & \\
					$(n_1,n_2)$ & $(0,n_2)$ & (1,1) & (1,2) & (1,3) & (1,4) \\
					\hline
					& & & & & \\
					$n^0_{(n_1,n_2)}$ & $-$276 &110646 & 6063822  &149000544 & 2918573208 \\
					\hline
					\hline
					& & & & & \\
					$(n_1,n_2)$ & $(r/3,0)$ & (1/3,3) & (1/3,6) & (1/3,9) & (1/3.12) \\
					\hline
					& & & & & \\
					$n^0_{(n_1,n_2)}$ & 378 & 110646& 6063822 & 164258118  & 2918573208 \\
					\hline
					\hline
					& & & & & \\
					$(n_1,n_2)$ & $(1/3,-1)$ & (1/3,1) & (4/3,1) & (7/3,1) & (10/3,1)  \\
					\hline
					& & & & & \\
					$n^0_{(n_1,n_2)}$ & 18 & 3066 & -463800 & 19292628 & 445321248 \\
					\hline
					\hline
					& & & & & \\
					$(n_1,n_2)$ & $(2/3,-1)$ & (2/3,1) & (5/3,1) & (8/3,1) & (11/3,1)  \\
					\hline
					& & & & &  \\
					$n^0_{(n_1,n_2)}$ & 6 & 9108&-1345986 & 51211656 & 1089299772 \\
					\hline
				\end{tabular}
		}}
	\end{center}
	\vspace{-0.5cm}
	\caption{$N=3$, 3A orbifold. }
	\renewcommand{\arraystretch}{0.5}
\end{table}

\begin{table}[H]
	\renewcommand{\arraystretch}{0.5}
	\begin{center}
		\vspace{0.5cm}
		\setlength{\doublerulesep}{20\arrayrulewidth}
		{\scriptsize{
				\begin{tabular}{|c|c|c|c|c|c|}
					\hline
					& & & & & \\
					$(n_1,n_2)$ & $(0,n_2)$ & (1,1) & (1,2) & (1,3) & (1,4) \\
					\hline
					& & & & & \\
					$ n^0_{(n_1,n_2)}$ &$-520$& 72750 & 1490560 & 103674750 & 1820321750 \\
					\hline
					\hline
					& & & & & \\
					$(n_1,n_2)$ & $(r/5,0)$ & (1/5,5) & (2/5,5) & (3/5,5) & (4/5,5) \\
					\hline
					& & & & &  \\
					$n^0_{(n_1,n_2)}$ & 250 & 72750& 3892000 & 103674750 & 1820321750 \\
					\hline
					\hline
					& & &  & &  \\
					$(n_1,n_2)$ &$(1/5,-1)$& (1/5,1) & (6/5,1) & (11/5,1) & (16/5,1)    \\ 
					\hline
					& & &  & &  \\
					$n^0_{(n_1,n_2)}$ & 14 & 1032 & 159648 & 7172574 & 175994068  \\
					\hline
					\hline
					& & &  & & \\
					$(n_1,n_2)$ &$(2/5,-1)$& (2/5,1) & (7/5,1) & (12/5,1) & (17/5,1)\\ 
					\hline
					& & &  & &   \\
					$n^0_{(n_1,n_2)}$ & 8 & 3040& 450704 & 17378724 & 375804020  \\
					\hline
					\hline
					& & &  & &   \\
					$(n_1,n_2)$ &$(3/5,-1)$ & (3/5,1) & (8/5,1) & (13/5,1) & (18/5,1) \\ 
					\hline
					& & &  \\
					$n^0_{(n_1,n_2)}$ & 6 & 4062& 605260 & 11893715 & 260500896  \\
					\hline
					\hline
					& & &  & &  \\
					$(n_1,n_2)$ & $(4/5,-1)$ & (4/5,1) & (9/5,1) & (14/5,1) & (19/5,1)  \\ 
					\hline
					& & & & &    \\
					$n^0_{(n_1,n_2)}$ & 2& 7082& 1045848 & 39672656 & 840212212  \\
					\hline
				\end{tabular}
		}}
	\end{center}
	\vspace{-0.5cm}
	\caption{$N=5$, 5A orbifold. }
	\renewcommand{\arraystretch}{-0.3}
\end{table}

\begin{table}[H]
	\renewcommand{\arraystretch}{0.5}
	
	\begin{center}
		\vspace{0.5cm}
		\setlength{\doublerulesep}{20\arrayrulewidth}
		{\scriptsize{
				\begin{tabular}{|c|c|c|c|c|c|}
					\hline
					& & & & & \\
					$(n_1,n_2)$ & (0,$n_2$) & (1,1) & (1,2) & (1,3) & (1,4)  \\
					\hline
					& & & & &  \\
					$n^0_{(n_1,n_2)}$ & $-$642 & 54309 & 2881360 & 76296885 & 1333671600 \\
					\hline
					\hline
					& & &  & & \\
					$(n_1,n_2)$ &(r/7,0) & (1/7,7) & (2/7,7) & (3/7,7) & (4/7,7) \\ 
					\hline
					& & &  & &  \\
					$n^0_{(n_1,n_2)}$ & 187 & 54309 & -2881360 & 76296885 & 1333671600  \\
					\hline
					\hline
					& & &  & &  \\
					$(n_1,n_2)$ &$(1/7,-1)$& (1/7,1) & (8/7,1) & (15/7,1) & (22/7,1) \\ 
					\hline
					& & &  & &   \\
					$n^0_{(n_1,n_2)}$ & 12 & 519 & 81348 & 3813632 & 96969840 \\
					\hline
					\hline
					& & &  & &  \\
					$(n_1,n_2)$ & $(2/7,-1)$ & (2/7,1) & (9/7,1) & (16/7,1) & (23/7,1)  \\ 
					\hline
					& & &  & &   \\
					$n^0_{(n_1,n_2)}$ & 6 & 1525& 227875 & 9059700  & 201601562 \\
					\hline
					\hline
					& & & & &    \\
					$(n_1,n_2)$ & $(3/7,-1)$& (3/7,1) & (10/7,1) & (17/7,1) & (24/7) \\ 
					\hline
					& & &  & &   \\
					$n^0_{(n_1,n_2)}$ & 7 & 2034 & 304146 & 12117044 & 268891664  \\
					\hline
					\hline
					& & &  & &  \\
					$(n_1,n_2)$ & $(4/7,-1)$ & (4/7,1) & (11/7,1) & (18/7,1) & (25/7,1)   \\ 
					\hline
					& & & & &   \\
					$n^0_{(n_1,n_2)}$ & 4 & 3540& 522435 & 19770135 & 418523073 \\
					\hline
					\hline
					& & & & &  \\
					$(n_1,n_2)$ & $(5/7,-1)$ & (5/7,1) & (12/7,1) & (19/7,1) & (26/7,1)   \\ 
					\hline
					& & &  & &   \\
					$n^0_{(n_1,n_2)}$ & 3 &3052& 456716  &9122877 & 202745998 \\
					\hline
					\hline
					& & & & &   \\
					$(n_1,n_2)$  &$(6/7,-1)$& (6/7,1) & (13/7,1) & (20/7,1) & (27/7,1)   \\ 
					\hline
					& & &  & &  \\
					$n^0_{(n_1,n_2)}$ &1& 6062& 892266 & 33394384 &697917622  \\
					\hline
				\end{tabular}
		}}
	\end{center}
	\vspace{-0.5cm}
	\caption{$N=7$, 7A orbifold. }
	\renewcommand{\arraystretch}{0.5}
\end{table}

\begin{table}[H]
	\renewcommand{\arraystretch}{0.5}
	\begin{center}
		\vspace{-0.5cm}
		\setlength{\doublerulesep}{20\arrayrulewidth}
		{\scriptsize{
				\begin{tabular}{|c|c|c|c|c|c|}
					\hline
					& & &  & &  \\
					$(n_1,n_2)$ & (0,$n_2$) & (1,1) & (1,2) & (1,3) & (1,4) \\ 
					\hline
					& & & & &   \\
					$ n^0_{(n_1,n_2)}$ &$-400$& 75776 &5323136 & 120004608 & 1941741952 \\
					\hline
					\hline
					& & & & &   \\
					$(n_1,n_2)$ & $(r/4,0),r=1,3$ & (1/4,4) & (1/4,8) & (1/4,12) & (1/4,16) \\
					\hline 
					& & & & &  \\
					$n^0_{(n_1,n_2)}$ & 256 & 75776 & 4335616 & 120004608 & 2156013568 \\
					\hline
					\hline
					& & & & &   \\
					$(n_1,n_2)$ &$(1/4,-1)$ &$(1/4,-2)$& (1/4,$-3$) & (1/4,1) & (5/4,1)  \\ 
					\hline
					& & & & &   \\
					$n^0_{(n_1,n_2)}$ & 16 & 8 & 4 & 2040 & 307204 \\
					\hline
					\hline
					& & & & &  \\
					$(n_1,n_2)$ &$(1/2,-1)$& (1/2,0) & (1/2,1)  & (1/2,2) & (3/2,1) \\ 
					\hline
					& & &  & &   \\
					$n^0_{(n_1,n_2)}$ & 8 & 368 & 4064 & 45952&  606288  \\
					\hline
					\hline
					& & &  & &   \\
					$(n_1,n_2)$ &$(3/4,-1)$& (3/4,1) & (7/4,1) & (11/4,1) & (15/4,1) \\ 
					\hline
					& & &  & &  \\
					$n^0_{(n_1,n_2)}$ & 4 & 8096 & 1196400 & 45505376 & 965922096  \\
					\hline
				\end{tabular}
		}}
	\end{center}
	\vspace{-0.5cm}
	\caption{$N=4$, 4B orbifold }
	\renewcommand{\arraystretch}{-0.3}
\end{table}

\begin{table}[H]
	\renewcommand{\arraystretch}{0.5}
	\begin{center}
		\vspace{0.5cm}
		\setlength{\doublerulesep}{20\arrayrulewidth}
		{\scriptsize{
				\begin{tabular}{|c|c|c|c|c|c|c|}
					\hline
					& & & & & & \\
					$(n_1,n_2)$ & (0,$n_2$) & (1,1) & (1,2) & (1,3) & (1,4) & (1,5) \\ 
					\hline
					& & &   & & & \\
					$ n^0_{(n_1,n_2)}$  &$-524$  & 39912 &  3596922  & 97260000 & 1599680040 & 17990607384 \\
					\hline
					\hline
					& & & & & &   \\
					$(n_1,n_2)$ & $(r/6,0),r=1,5$ & (1/6,6) & (1/6,12) & (1/6,18) & (1/6,24) & (1/6,30)  \\ 
					\hline
					& & & & & &  \\
					$n^0_{(n_1,n_2)}$ & 132 & 39912 & 2466900 & 71374608 & 1318893168 & 17990607384 \\
					\hline
					\hline
					& & & & & &   \\
					$(n_1,n_2)$ & $(r/3,0),r=1,2$ & (1/2,0) & (1/3,3) & (1/2,2) & (1/3,6) & (1/2,4) \\ 
					\hline
					& & & & & &  \\
					$n^0_{(n_1,n_2)}$ & 246 & 248& 70734& 71728& 3596922 & 3737432 \\
					\hline
					\hline
					& & &  & & &  \\
					$(n_1,n_2)$ &$(1/6,-1)$& (5/6, $-1$)&(1/6,1) & (5/6,1) & (7/6,1) & (11/6,1) \\
					\hline
					& & & & & &   \\
					$n^0_{(n_1,n_2)}$ & 12 & 2 & 1024 & 6072 & 155620 & 897300 \\
					\hline
					\hline
					& & & & & &  \\
					$(n_1,n_2)$ &$(1/6,-2)$& (1/6,$-3$) & (1/6,$-4$) &$(1/2,-1)$ & (1/2,1) & (3/2,1) \\ 
					\hline
					& & &   & & &  \\
					$n^0_{(n_1,n_2)}$ & 8 & 6 & 4 & 6 & 3054 & 457740   \\
					\hline
					\hline
					
					& & & & & &  \\
					$(n_1,n_2)$ &$(1/3,-1)$& (2/3,$-1$) & (1/3,1) & (2/3,1) & (4/3,1) & (5/3,1) \\ 
					\hline
					& & & & & &  \\
					$n^0_{(n_1,n_2)}$ & 8 & 2032 & 303152 & 4 & 4056 & 602240   \\
					\hline
				\end{tabular}
		}}
	\end{center}
	\vspace{-0.5cm}
	\caption{$N=6$, 6A orbifold }
	\renewcommand{\arraystretch}{-0.3}
\end{table}

\begin{table}[H]
	\renewcommand{\arraystretch}{0.5}
	\begin{center}
		\vspace{0.5cm}
		\setlength{\doublerulesep}{20\arrayrulewidth}
		{\scriptsize{
				\begin{tabular}{|c|c|c|c|c|c|c|}
					\hline
					& & & & & & \\
					$(n_1,n_2)$ & (0,$n_2$) & (1,1) & (1,2) & (1,3) & (1,4) & (1,5)  \\ 
					\hline
					& & & & & &  \\
					$ n^0_{(n_1,n_2)}$  &$-644$ & 37888 & 2661568 & 60002304 & 1452283424 & 14426028032 \\
					\hline
					\hline
					& & & & & & \\
					$(n_1,n_2)$ & $(r/8,0),r={\rm odd}$ & (1/8,8) & (1/8,16) & $(r/4,0),r={\rm odd}$ & (1/4,4) & (1/4,8)   \\ 
					\hline
					& & & & & & \\
					$n^0_{(n_1,n_2)}$ & 128 & 37888 & 2167808  & 184 & 22976 & 2661568 \\
					\hline
					\hline
					& & & & & & \\
					$(n_1,n_2)$ & $(1/4,-2)$& (1/4,$-3$) & (1/2,$-1$) & $(1/2,0)$ & (1/2,2) & (1/2,4)  \\ 
					\hline
					& & & & & & \\
					$n^0_{(n_1,n_2)}$ &  6 & 2 & 4 & 244 & 22976 & 1880800 \\
					\hline
					\hline
					& & & & & &  \\
					$(n_1,n_2)$ &$(1/8,-1)$& (1/8,1) & (3/8,1) & (5,8,1) & (7/8,1) & (9/8,1)  \\
					\hline
					& & & & & &  \\
					$n^0_{(n_1,n_2)}$ & 8 & 517 & 2028 & 3038 & 4056 & 80334 \\
					\hline
					\hline
					& & & & & & \\
					$(n_1,n_2)$ &$(1/8,-2)$& (1/8,$-3$) & (1/8,$-4$) &$(1/8,-5)$& (1/8,$-6$) & (1/8,$-7$)\\ 
					\hline
					& & & & & &  \\
					$n^0_{(n_1,n_2)}$ & 8 & 6 & 4 & 4 & 2 & 1    \\
					\hline
					\hline
					& & & & & & \\
					$(n_1,n_2)$ &$(1/4,-1)$ &$(3/4,-1)$& (1/4,1) & (3/4,1) & (5/4,1) & (7/4,1)  \\ 
					\hline
					& & & & & & \\
					$n^0_{(n_1,n_2)}$ & 16 & 2 & 1020 & 8096 & 153602 & 1196400   \\
					\hline
				\end{tabular}
		}}
	\end{center}
	\vspace{-0.5cm}
	\caption{$N=8$, 8A orbifold }
	\renewcommand{\arraystretch}{-0.3}
\end{table}

\section{List of twisted elliptic genera} \label{frslist}

In this appendix we  provide the list of the twisted elliptic genera used to determine the $f^{(r,s)}$
defined in (\ref{defcrsfrs})  in this paper.  
We call the classes $2A, 3A, 5A, 7A, 4B, 6A, 7A$, CHL orbifolds. 
When $g'$ corresponding to these classes act on $K3$, it can be seen that 
the values of the Hodge number $h_{1,1}\geq 1$. 
The twisted elliptic genus of  classes $2B$, $3B$ is such that 
their twining character coincides with that 
constructed in \cite{Cheng:2010pq,Eguchi:2010fg,Gaberdiel:2010ch}. 
But they do not correspond to 
$M_{24}$ symmetry. In fact in $M_{24}$, the classes have order $2$ and $3$, 
however the orbifolds we  construct have order $4$ and $9$. 
The $2B$ twisted elliptic genus has been  constructed 
in \cite{Gaberdiel:2013psa} by considering $K3$ as 6, $SU(2)$ WZW models at level 1. 
While $3B$ twisted elliptic genus we give here was constructed in \cite{Chattopadhyaya:2017ews}. 
\subsection*{ Conjugacy class $ pA$,  $p$ =\{1,2,3,5,7\}}
\bea
F^{(0, 0)} &=& = \frac{8}{p} A(\tau, z), \\ \nonumber 
F^{(0,s)}&=&\frac{8}{p(p+1)}A(\tau, z)-B (\tau, z) 
\left(\frac{2}{p+1}{\cal E}_{p}(\tau)\right)\\ \nn
F^{(r, rs ) }&=&\frac{8}{p(p+1)}A(\tau, z) +B(\tau, z) 
\left(\frac{2}{p(p+1)}{\cal E}_{p}(\frac{\tau+s}{p})\right)
\eea

\subsection*{ Conjugacy class $11A$}

Going through these steps we obtain the following formula for the 
twisted elliptic genus for $11A$. 
\begin{eqnarray}\label{answ11A}
F^{(0, 0)} &=& = \frac{8}{11} A(\tau, z), \\ \nonumber 
F^{(0,s)}&=&\frac{2}{33}A(\tau, z)-B (\tau, z) 
\left(\frac{1}{6}{\cal E}_{11}(\tau) -\frac{2}{5}\eta^2(\tau)\eta^2(11\tau)\right) ,\\ \nn
F^{(r, rs ) }&=&\frac{2}{33}A(\tau, z) +B(\tau, z) 
\left(\frac{1}{66}{\cal E}_{11}(\frac{\tau+s}{11})-\frac{2}{55}
\eta^2(\tau+s)\eta^2(\frac{\tau+s}{11})\right) .
\end{eqnarray}

\subsection*{Conjugacy class $23A/B$}
\bea \label{23abtwist}
F^{(0,k)}(\tau, z) &=&\frac{1}{23}\left(\frac{1}{3}A-B\left(\frac{23}{12}{\cal E}_{23}(\tau) -\frac{23}{22}f_{23,1}(\tau)-\frac{161}{22}\eta^2(\tau)\eta^2(23\tau)\right)\right) ,\\ \nn
F^{(r, rk) }(\tau, z) &=&\frac{1}{23}
\left[\frac{1}{3}A+B\left(\frac{1}{12}{\cal E}_{23}(\frac{\tau+k}{23})-
\frac{1}{22}f_{23,1}(\frac{\tau+k}{23})-\frac{7}{22}\eta^2(\tau+k)\eta^2(\frac{\tau+k}{23})\right)\right].
\eea
where
\begin{equation}\label{f321}
f_{23,1}(\tau)=2\frac{\eta^3(\tau)\eta^3(23\tau)}{\eta(2\tau)\eta(46\tau)}+8\eta(\tau)\eta(2\tau)\eta(23\tau)\eta(46\tau)+8 \eta^2(2\tau)\eta^2(46\tau)+5\eta^2(\tau)\eta^2(23\tau)
\end{equation}

\subsection*{Conjugacy class $4B$}

The twisted elliptic genus for this class is given by 
\bea \label{twch4b}
F^{(0,0)}(\tau, z) &=&2A(\tau, z) ,\nn \\ 
F^{(0,1)}(\tau, z) &=& F^{(0,3)}(\tau, z) =
\frac{1}{4}\left[\frac{4A}{3}-B \left(-\frac{1}{3}{\cal E}_2(\tau)+2{\cal E}_4(\tau) \right)\right], \\ \nn
F^{(1,s)}(\tau, z) &=& F^{(3,3s)}=\frac{1}{4}\left[\frac{4A}{3}+
B \left(-\frac{1}{6}{\cal E}_2(\frac{\tau+s}{2})+\frac{1}{2}{\cal E}_4(\frac{\tau+s}{4}) \right)\right], \\ \nn
F^{(2,1)}(\tau, z) &=& F^{(2,3)}=\frac{1}{4}\left(\frac{4A}{3}
-\frac{B}{3} (5{\cal E}_2(\tau)-6{\cal E}_4(\tau)\right) ,\\ \nn
F^{(0,2)}(\tau, z) &=&\frac{1}{4}\left(\frac{8A}{3}-\frac{4B}{3}{\cal E}_2(\tau)\right) ,\\ \nn
F^{(2,2s)}(\tau, z) &=&\frac{1}{4}\left(\frac{8A}{3}+\frac{2B}{3}{\cal E}_2(\frac{\tau+s}{2})\right) .
\eea

\subsection*{Conjugacy class $6A$}

The twisted elliptic 
genus for $6A$ are given by 
\bea\label{6a1}
F^{(0,0)}&=&\frac{4}{3}A;\quad  F^{(0,1)}=  F^{(0,5)};\quad  F^{(0,2)}=  F^{(0,4)};\\ \nn
F^{(0,1)} &=&\frac{1}{6}\left[\frac{2A}{3}-B \left(-\frac{1}{6}{\cal E}_2(\tau)
-\frac{1}{2}{\cal E}_3(\tau)+\frac{5}{2}{\cal E}_6(\tau) \right)\right],\\ \nn
F^{(0,2)}&=& \frac{1}{6}\left[2A-\frac{3}{2}B {\cal E}_3(\tau)\right], \\ \nn
F^{(0,3)}&=& \frac{1}{6}\left[\frac{8A}{3}-\frac{4}{3}B {\cal E}_2(\tau)\right].
\eea
\bea \label{6a2} \nn
F^{(1, k)}&=&F^{(5, 5k)}=\frac{1}{6}\left[\frac{2A}{3}+B \left(-\frac{1}{12}{\cal E}_2(\frac{\tau+k}{2})-\frac{1}{6}{\cal E}_3(\frac{\tau+k}{3})+\frac{5}{12}{\cal E}_6(\frac{\tau+k}{6}) \right)\right], \\
\eea
\bea\label{6a3}
F^{(2, 2k+1)}&=& \frac{A}{9}+\frac{B}{36}\left[ {\cal E}_3(\frac{\tau+2+k}{3})
+ {\cal E} _2(\tau) - {\cal E}_2( \frac{\tau  +k +2}{3} ) \right], \\ \nn
F^{(4, 4k+1)}&=& \frac{A}{9}+\frac{B}{36}\left[{\cal E}_3(\frac{\tau+1+k}{3})
+ {\cal E} _2(\tau) - {\cal E}_2( \frac{\tau  +k +1}{3} ) \right], \\ \nn
F^{(3, 1)}&=&F^{(3, 5)}= \frac{A}{9}-\frac{B}{12}{\cal E}_3(\tau)-\frac{B}{72}{\cal E}_2(\frac{\tau+1}{2})+\frac{B}{8}{\cal E}_2(\frac{3\tau+1}{2}), \\ \nn
F^{(3, 2)}&=&F^{(3, 4)}= \frac{A}{9}-\frac{B}{12}{\cal E}_3(\tau)-\frac{B}{72}{\cal E}_2(\frac{\tau}{2})+\frac{B}{8}{\cal E}_2(\frac{3\tau}{2}), \\ \nn
\eea
\bea\label{6a4}
F^{(2r,2rk)}&=& \frac{1}{6}\left[2A+\frac{1}{2}B {\cal E}_3(\frac{\tau+k}{3})\right],  \nn \\
F^{(3,3k)}&=& \frac{1}{6}\left[\frac{8A}{3}+\frac{2}{3}B {\cal E}_2(\frac{\tau+k}{2})\right].
\eea


\subsection*{Conjugacy class $8A$}

\bea
F^{(0, 0)} (\tau, z) &=& A(\tau, z), \\ \nonumber
F^{(0,1)}&=& F^{(0,3)}= F^{(0,5)}= F^{(0,7)},\\ \nn
&=&\frac{1}{8}
\left[\frac{2A}{3}-B \left(-\frac{1}{2}{\cal E}_4(\tau)+\frac{7}{3}{\cal E}_8(\tau) \right)\right].
\eea

\be
F^{(r, rk)}(\tau, z) =\frac{1}{8}\left[\frac{2A}{3}+\frac{B}{8} \left(-{\cal E}_4(\frac{\tau+k}{4})+\frac{7}{3}{\cal E}_8(\frac{\tau+k}{8})\right)\right].
\ee
where $r=1,3,5,7$.

\bea
F^{(2,1)}&=& F^{(6,3)}= F^{(2,5)}= F^{(6,7)},\\ \nn
&=& \frac{1}{8}\left[\frac{2A}{3}+\frac{B}{3} \left(-{\cal E}_2(2\tau)+\frac{3}{2}{\cal E}_4(\frac{2\tau+1}{4}) \right)\right];\\ \nn
F^{(2,3)}&=& F^{(6,5)}= F^{(2,7)}= F^{(6,1)},\\ \nn
&=& \frac{1}{8}\left[\frac{2A}{3}+\frac{B}{3} \left(-{\cal E}_2(2\tau)+\frac{3}{2}{\cal E}_4(\frac{2\tau+3}{4}) \right)\right].
\eea
\bea
F^{(0,2)}&=& F^{(0,6)}=\frac{1}{8}\left(\frac{4A}{3}-B \left(-\frac{1}{3}{\cal E}_2(\tau)+2{\cal E}_4(\tau) \right)\right),\\ \nn
F^{(0,4)}&=&\frac{1}{8}\left(\frac{8A}{3}-\frac{4B}{3}{\cal E}_2(\tau)\right),\\ \nn
F^{(2,2s)}&=& F^{(6,6s)}=\frac{1}{8}\left(\frac{4A}{3}+B \left(-\frac{1}{6}{\cal E}_2(\frac{\tau+s}{2})+\frac{1}{2}{\cal E}_4(\frac{\tau+s}{4}) \right)\right),\\ \nn
F^{(4,4s)}&=&\frac{1}{8}\left(\frac{8A}{3}+\frac{2B}{3}{\cal E}_2(\frac{\tau+s}{2})\right),\\ \nn
F^{(4,2)}&=& F^{(4,6)}=\frac{1}{8}\left(\frac{4A}{3}-\frac{B}{3} (3{\cal E}_2(\tau)-4{\cal E}_2(2\tau)\right),\\ \nn
F^{(4,2k+1)}&=&\frac{1}{8}\left(\frac{2A}{3}+B \left(\frac{4}{3}{\cal E}_2(4\tau)-\frac{2}{3}{\cal E}_2(2\tau)-\frac{1}{2}{\cal E}_4(\tau) \right)\right).
\eea

%

\subsection*{ Conjugacy class $14A$}
\bea
F^{(0,1)}(\tau, z) = F^{(0,3)}= F^{(0,5)}= F^{(0,9)}= F^{(0,11)}= F^{(0,13)};\\ \nn
=\frac{1}{14}\left[\frac{A}{3}-B \left(-\frac{1}{36}{\cal E}_2(\tau)-\frac{7}{12}{\cal E}_7(\tau)+\frac{91}{36}{\cal E}_{14}(\tau)\right. \right. \\ \nn
\left.\left. -\frac{14}{3} \eta(\tau)\eta(2\tau)\eta(7\tau) \eta(14\tau)\right) \right];
\eea
\bea
F^{(r, rk)}&=&\frac{1}{14}\left[\frac{A}{3}+B \left(-\frac{1}{72}{\cal E}_2(\frac{\tau+k}{2})-\frac{1}{12}{\cal E}_7(\frac{\tau+k}{7})+\frac{13}{72}{\cal E}_{14}(\frac{\tau+k}{14}) \right. \right. \\ \nn
&&\left.\left. -\frac{1}{3} \eta(\tau+k)\eta(\frac{\tau+k}{2})\eta(\frac{\tau+k}{7}) \eta(\frac{\tau+k}{14})
\right) \right];
\eea
where $r$=1,3,5,9,11,13 and $rk$ is Mod 14.

\bea
F^{(2r, 2rk+7)}&=&\frac{1}{14}\left[\frac{A}{3}+B \left(-\frac{1}{6}{\cal E}_2(\tau)-\frac{1}{12}{\cal E}_7(\frac{\tau+k}{7})+\frac{1}{3}{\cal E}_{7}(\frac{2\tau+2k}{7}) \right. \right. \\ \nn
&&\left.\left. -\frac{2}{3} \eta(\tau+k)\eta(2\tau+2k)\eta(\frac{\tau+k}{7}) \eta(\frac{2\tau+2k}{7})\right) \right];
\eea
where $k$ runs from 0 to 6 and except 3 and $r$ from 1 to 6.
\bea
F^{(7,2k+1)}&=&\frac{1}{14}\left[\frac{A}{3}+B \left(-\frac{7}{12}{\cal E}_7(\tau)+\frac{49}{72}{\cal E}_2(\frac{7\tau+1}{2})-\frac{1}{72}{\cal E}_2(\frac{\tau+1}{2}) \right. \right. \\ \nn
&&\left.\left. +\frac{7}{3} 
e^{i\pi 11/12}\eta(\tau)\eta(7\tau)\eta(\frac{\tau+1}{2}) \eta(\frac{7\tau+1}{2})\right) \right];\\ \nn
F^{(7,2k)}&=&\frac{1}{14}\left[\frac{A}{3}+B \left(-\frac{7}{12}{\cal E}_7(\tau)+\frac{49}{72}{\cal E}_2(\frac{7\tau}{2})-\frac{1}{72}{\cal E}_2(\frac{\tau}{2}) \right. \right. \\ \nn
&&\left.\left. +\frac{7}{3} \eta(\tau)\eta(7\tau)\eta(\frac{\tau}{2}) \eta(\frac{7\tau}{2})\right) \right].
\eea
\bea
F^{(0,0)}&=&\frac{4}{7}A.\\ 
F^{(0,2k)}&=&\frac{1}{14}\left[A-\frac{7}{4}B{\cal E}_7(\tau)\right] \quad k\; {\rm runs\;from} \;1\;{\rm to}\;6 ,\\ \nn
F^{(2r,2rk)}&=&\frac{1}{14}\left[A+\frac{1}{4}B{\cal E}_7(\frac{\tau+k}{7})\right]; \quad k\; {\rm runs\;from} \;0\;{\rm to}\;6. \\ 
F^{(0,7)}&=&\frac{1}{14}\left[\frac{8}{3}A-\frac{4}{3}B{\cal E}_2(\tau)\right] , \\ \nn
F^{(7,7k)}&=&\frac{1}{14}\left[
\frac{8}{3}A+\frac{2}{3}B{\cal E}_2(\frac{\tau+k}{2})\right] \quad k\; {\rm runs\;from} \;0\;{\rm to}\;1.
\eea

\subsection*{Conjugacy class $15A$}
\bea
&& F^{(0,1)}(\tau, z) = F^{(0,2)}= F^{(0,4)}= F^{(0,7)}= F^{(0,8)}= F^{(0,11)}= F^{(0,13)}=F^{(0,14)};\\ \nn
&&=\frac{1}{15}\left[
\frac{A}{3}-B \left(-\frac{1}{16}{\cal E}_3(\tau)-\frac{5}{24}{\cal E}_5(\tau)+\frac{35}{16}
{\cal E}_{15}(\tau) -\frac{15}{4} \eta(\tau)\eta(3\tau)\eta(5\tau) \eta(15\tau)\right) \right].
\eea
\bea
F^{(r, rk)}&=&\frac{1}{15}
\left[\frac{A}{3}+B \left(-\frac{1}{48}{\cal E}_3(\frac{\tau+k}{3})-\frac{1}{24}{\cal E}_5(\frac{\tau+k}{5})+\frac{7}{48}{\cal E}_{15}(\frac{\tau+k}{15}) \right. \right. \\ \nn
&&\left.\left. -\frac{1}{4} \eta(\tau+k)\eta(\frac{\tau+k}{3})\eta(\frac{\tau+k}{5}) 
\eta(\frac{\tau+k}{15})\right) \right];
\eea
where $r$=1,2,4,7,8,11,13,14 and $rk$ is mod 15.
The sectors belonging to the $5A$ and $3A$ sub-orbits are given by 
\bea
F^{(0,0)}&=&\frac{8}{15}A.\\ 
F^{(0,3k)}&=&\frac{1}{15}\left(\frac{4}{3}A-\frac{5}{3}B{\cal E}_5(\tau)\right) \quad k\; {\rm runs\;from} \;1\;{\rm to}\;4;\\ \nn
F^{(3r,3rk)}&=&\frac{1}{15}\left(\frac{4}{3}A+\frac{1}{3}B{\cal E}_5(\frac{\tau+k}{5})\right); \quad k\; {\rm runs\;from} \;0\;{\rm to}\;4. \\ 
F^{(0,5k)}&=&\frac{1}{15}\left(2A-\frac{3}{2}B{\cal E}_3(\tau)\right) ;\\ \nn
F^{(5r,5rk)}&=&\frac{1}{15}\left(2A+\frac{1}{2}B{\cal E}_3(\frac{\tau+k}{3})\right) \quad k\; {\rm runs\;from} \;0\;{\rm to}\;2.
\eea
Finally the remaining sectors are given by 
\bea
F^{(3r, 5+3rk)}&=&\frac{1}{15}\left(\frac{A}{3}+B \left(-\frac{1}{4}{\cal E}_3(\tau)-\frac{1}{24}{\cal E}_5(\frac{\tau+k}{5})+\frac{3}{8}{\cal E}_{5}(\frac{3\tau+3k}{5}) \right. \right. \\ \nn
&&\left.\left. -\frac{3}{4} \eta(\tau+k)\eta(3\tau+3k)\eta(\frac{\tau+k}{5}) \eta(\frac{3\tau+3k}{5})\right) \right), \\ \nonumber
F^{(3r, 10+3rk)}&=&\frac{1}{15}\left(\frac{A}{3}+B \left(-\frac{1}{4}{\cal E}_3(\tau)-\frac{1}{24}{\cal E}_5(\frac{\tau+k}{5})+\frac{3}{8}{\cal E}_{5}(\frac{3\tau+3k}{5}) \right. \right. \\ \nn
&&\left.\left. -\frac{3}{4}e^{ -\frac{2\pi i}{5}} 
\eta(\tau+k)\eta(3\tau+3k)\eta(\frac{\tau+k}{5}) \eta(\frac{3\tau+3k}{5})\right) \right);
\eea
where $k$ runs from 0 to 4 and $s$=1 to 4.
\bea
F^{(5r, 3s+5rk)}&=&\frac{1}{15}\left(\frac{A}{3}+B \left(\frac{5}{24}{\cal E}_5(\tau)+\frac{1}{12}{\cal E}_3(\frac{\tau+k}{3})-\frac{5}{24}{\cal E}_{5}(\frac{\tau+k}{3}) \right. \right. \\ \nn
&&\left.\left. +\frac{5}{4} \eta(\tau+k)\eta(5\tau+5k)\eta(\frac{\tau+k}{3}) \eta(\frac{5\tau+5k}{3})\right) \right);
\eea
where $k$ runs from 0 to 2 and $s$=1 to 2.

\subsection*{Conjugacy class 2B}
\bea\label{comp2b}
F^{(0,0)}(\tau, z) &=&2A;\quad  F^{(0,1)}(\tau, z) =  F^{(0,3)}(\tau, z), \\ \nn
F^{(0,1)}(\tau, z) &=&\frac{B(\tau, z) }{2}({\cal E}_2(\tau)-{\cal E}_4(\tau)),\\ \nn
F^{(0,2)}(\tau, z)  &=&-\frac{2A(\tau, z) }{3}-\frac{2B(\tau, z) }{3}{\cal E}_2(\tau) ,\\ \nn
F^{(1,s)}(\tau, z) &=& F^{(3,3s)}=-\frac{B(\tau, z) }{4}({\cal E}_2(\frac{\tau+s}{2})-
{\cal E}_4(\frac{\tau+s}{4})),\\ \nn
F^{(2,1)}(\tau, z)  &=&F^{(2,3)}=\frac{B(\tau, z) }{2}(-\frac{1}{6}{\cal E}_2(\tau)+
\frac{2}{3}{\cal E}_2(2\tau)),\\ \nn
F^{(2,2s)}(\tau, z)  &=&-\frac{2A(\tau, z) }{3}+\frac{B(\tau, z) }{3}{\cal E}_2(\frac{\tau+s}{2}) . \\ \nn
\eea
\subsection*{Conjugacy class 3B}
\bea
F^{(0,0)} (\tau, z) &=&\frac{8A(\tau, z) }{9},\\ \nonumber 
F^{(0,1)} (\tau, z) &=&  F^{(0,2)}= F^{(0,4)}= F^{(0,5)}= F^{(0,7)}= F^{(0,8)};\\ \nn
F^{(0,1)}(\tau, z)  &=&-\frac{2B(\tau, z) }{9}\frac{\eta^6(\tau)}{\eta^2(3\tau)},\\ \nn
F^{(0,3)}(\tau, z)  &=&-\frac{A(\tau, z) }{9}-\frac{B(\tau, z) }{4}{\cal E}_3(\tau) ,\\ \nn
F^{(r,rs)}(\tau, z)  &=&\frac{2B(\tau, z) }{3}\frac{\eta^6(\tau+s)}{\eta^2(\frac{\tau+s}{3})}, \quad\quad r=1,2,4,5,7,8 \\ \nn
F^{(3,1)}(\tau, z)  &=&-\frac{2B(\tau, z)}{9}e^{2\pi i/3}\frac{\eta^6(\tau)}{\eta^2(3\tau)} ,\\ \nn
&=&  F^{(3,4)}=F^{(3,7)}= F^{(6,2)}=F^{(6,8)}=F^{(6,5)};\\ \nn
F^{(3,2)}(\tau, z)  &=&-\frac{2B(\tau, z) }{9}e^{4\pi i/3}\frac{\eta^6(\tau)}{\eta^2(3\tau)},
\\ \nn
&=& F^{(3,5)}=F^{(3,8)} = F^{(6,1)}=F^{(6,7)}=F^{(6,4)}; \\ \nn
F^{(3r,3rk)}(\tau, z)  &=&-\frac{A(\tau, z) }{9}+\frac{B(\tau, z) }{12}{\cal E}_3(\frac{\tau+k}{3}) .\\ \nn
\eea

\providecommand{\href}[2]{#2}\begingroup\raggedright\endgroup


\end{document}